# A Smart Chair for Health Monitoring in Daily Life

[z]Nguyen Thi Minh Huong[1], *Vo Quoc Bao[1], *Nguyen Trung Hau[1], Huynh Quang Linh[1]
[1]Ho Chi Minh City University of Technology (HCMUT), VNU-HCM, Ho Chi Minh City, Viet Nam
Email: nguyentmhuong@hcmut.edu.vn

**Abstracts**

Recent research has focused on the risks associated with poor sitting posture and the impact of sitting on biological parameters, such as heart rate because prolonged sitting is common across all ages and professions. In this work, we propose a novel approach that can display simultaneously posture and heart rate in real-time. In this device, pressure sensors are embedded into a flexible separate cushion that is easily put on any chair to provide sitting behaviors, and a smartwatch-like PPG module is worn on the user's wrist. Regarding posture classification, pressure figures of ten pressure sensors under the seat bottom are inputs of four machine learning models, giving a high accuracy of 99 per cent. Besides, the electrocardiography (ECG) recording module is illustrated with the same result as a commercial device called DFRobot. One more advantage of this smart chair is it not only displays both sitting postures and heart rates simultaneously on external pieces of devices like laptops, mobile phones, or televisions through microcontrollers but also offers the relationship between them to help people adjust their sitting behaviours, avoiding to influence heart rate. The smart chair is expected to be useful equipment for sedentary- lifestyle people, especially for office workers.

*Keywords: Smart Chair, Electrocardiography, posture, pressure sensor, Photoplethysmography*

## 1. Introduction

The sedentary lifestyle has been a public concern in recent years. The statistics show that, on average, the population spends most of their daily time in a seated position, inevitably worsening their health, especially concerning bone and muscle issues [1]. While prolonged sitting is a mild problem because of support from the cutting-edge alarm system, the sitting posture receives more attention since the posture alarm devices are not completely effective. The ongoing research has proved that sitting posture is considered one of the major risk factors for an increasing low back pain (LBP) trend [2, 3]. The prevalence of low back pain has been affecting the economies of industrialized countries in many ways [4, 5] and occupations that require workers to sit for a large portion of the working day occupy a higher percentage of employees having LBP [6, 7]. Some studies also showed that people do not pay attention to their posture [8] and they tend to slouch rather than sit upright which is the best of all postures [9, 10]. Although plenty of already commercial smart chairs incorporate many useful features for users such as adjustable backrests and protectors, removable cushions, and so on, users still ignore them out of their habits. Therefore, a monitoring system integrated into a chair is an urgent solution to help users adjust their sitting postures, preventing several associated diseases. Many works and real products already exist and some efforts are being made to bring the most convenient chair to users. Patrick Vermander et.al. [11] present a relatively comprehensive overview of existing sitting posture monitoring systems in the literature. Accordingly, three popular systems are being used, camera-based systems [12, 13, 14, 15, 16, 17, 18], wearable devices [19, 20, 21, 22, 23, 24, 25], and systems of sensors located on the surfaces of assistance or support devices [26, 27, 28, 29, 30]. The cameras of the first system provide undesired personal information when they capture users' sitting pictures and inseparable difficulties in setting measurement conditions like light conditions and impediments between the camera and the subject and so on. This is the main reason for disregarding it in many works [11]. The second system shows the ability to collect data continuously and own a low-cost solution. However, it has many drawbacks, such as discomfort and loss of aesthetics, when users have to wear it. Furthermore, it is prone to moving due to rubbing against the users' clothes and movement, leading to a lack of signal stability [11]. They are the main reasons for the last to prevail. In this system, some factors should be discussed including the type of sensors, placement of sensors, experiment design and classifiers. Some sensors are used in the system, like textile pressure sensors [31, 32, 33], force sensing/sensitive sensors (FSR) [26, 34, 35, 36, 37], load cells [38, 39], and accelerometer sensors [40, 41]. Among them, the FSR is used the most because it does not require complicated electronic circuits, and is cost-effective [11, 42] although the others are still implemented in some chairs. This is one of the root reasons for our work to choose FSR. In addition to sensor selection, the arrangement of sensors on smart chairs is important because changing sitting postures only causes a subtle change in physical quantities that sensors are capable of sensing and sending signals to the classifier. In the literature, there are two common ways to distribute sensors, out of which places sensors only on the cushion and the other places sensors on both the cushion and the backrest. Placing sensors on the armrest is not popular. It seems more common for sensors to be placed on both the seat and the backrest because this is more efficient than the other positions [30, 42]. This has been not officially reported in the literature yet. However, this trend usually requires different types of sensors like capacitor sensors, acceleration sensors, infrared sensors and temperature sensors. With the supplement of sensors besides pressure sensors, smart chairs lose their portability because the other sensors are mounted on the backrest of the chair. We aim to design a smart cushion that monitors sitting postures and can be brought anywhere users want. Another solution to movable cushions is the pressure mats on the seat [32, 43, 44, 45, 46, 47, 48, 49, 59]. But it is expensive and computationally inefficient. To design a smart chair that can properly classify the sitting postures at the right cost and mobility, we place ten pressure sensors beneath a separate mat. To achieve good results, we also pay attention to distributing sensors suitably and have an effective experiment design to collect good training data, classifying in real-time properly in return.

One public concern is that cardiovascular diseases are increasing significantly in younger adults. There are plenty of devices that support users to monitor their heart rate like commercialized smartwatches [50] or others available in stores [51]. However, these devices only provide instant information about the heart rate and users only pay attention in specific situations like sick or exercising. Continuous monitoring of heart rate during sitting is necessary because they can review their habits in real-time, especially those who sit for a long time. Some smart chairs combined with heart rate modules released their achievements [39, 52]. However, their accuracy is a concern. In this work, we design a smartwatch-like heart rate measurement module using the Photoplethysmography (PPG) technique.

In general, there are multiple ways to design a smart chair that can effectively classify sitting postures. However, towards convenience, portability and cost savings, the main research contributions of this paper are: (1) This work proposes a system that requires a smaller number of sensors compared to previously published research but can capture minor changes in users' postures. This reduces the complexity of the system in terms of hardware structure, and classification model, prolonging the device's battery life and reducing cost. (2) By using only pressure sensors without different types of sensors, we produce a movable cushion that is easily put on any surface of an office chair. (3) Along with using fewer sensors, the symmetrical and suitable placement of sensors under the cushion yields positive practical results. (4) Classification results are significantly higher than previous related research as demonstrated in the discussion part. (5) Besides displaying sitting posture and heart rate simultaneously, some utilities that smart cushion and the heart rate module bring to users are reviewing and doing statistics about sitting postures and heart rate.

The rest of the paper is organized as follows. Section 2 describes methods applied in the paper, including two modules, a sitting posture classification module and a heart rate measuring module. Section 3 presents our experimental evaluation results. Section 4 discusses the performance of the proposed approach and compares it with other systems in the literature. Finally, Section 5 highlights the main conclusions and future work perspectives.

## 2. Methods

### *2.1 Sitting posture classification Module*

In this study, sitting posture classification is based on the pressure distribution of the system of ten sensors FSR (Force Sensing Resistor) under the wrapper of a separate seat cushion. The module uses the pressure sensor typed FSR-402 because each can properly operate with a measurement range of ten kilograms that fits the user's maximum weight of one hundred kilograms. Besides, its tiny size also brings some advantages. Firstly, it is easy to put many sensors on the surface without causing discomfort to users. Secondly, it brings a good performance because these sensors do not influence each other.

#### *2.1.1 Sensor position*

The sensor position significantly affects the accuracy of classification because improper placement of sensors may cause an overlap or a lack of output signals that can not help distinguish different sitting postures. In contrast, the placement of sensors is suitable, values of sensors can illustrate minor changes in a sitting position. The idea to arrange sensors was derived from the research on pressure maps when sitting by Lee et al. [53]. To be more specific, ten sensors are symmetrically placed about the axis of the rectangle-shaped cushion 45 cm in length and 35 cm in width as Figure 1. The pressure sensors are positioned on the surface of the rubber mat inside of the separate seat cushion instead of putting them on the surface like in some other research to prevent signal deviation due to the motion of sensors when the user sits on it and to enhance aesthetic value as well as not cause discomfort for users.

Figure 2 shows a real image of sensors on the mattress covered by the seat cushion. The seat cushion where the sensor system operates is separate, so it can be convenient to use everywhere by putting it on the surface of a smart chair.

#### *2.1.2 Acquisition and Transmission Module*

Some electrical components being used in this system include connection ports (connecting pressure sensors and voltage dividers), voltage dividers, buffer amplifiers, microprocessors, and screens. Each force sensor has 2 pins, one pin will be connected to the positive source, and the other will be connected to the voltage divider circuit. The connection port has 6 pins of which one will be connected to the positive power supplier while the others will connect to 5 pins of 5 sensors. Each connection port will be responsible for recording signals from one side of the chair, so 2 connections are needed to perform the task. The principle diagram is shown in Figure 3 plotted by Easy EDA software. To optimize space, we use adhesive electrical components and design a two-layer circuit, with the top side including the signal connection port from the sensor and voltage divider circuit, the buffer circuit IC LM324 will be placed on the bottom side. The pins of the microcontroller and the circuit of the OLED display can be connected on both sides. We use ten force sensors, so we will need ten resistors to form a voltage divider circuit, similarly with ten buffer circuits, we will need three LM324 ICs (each IC can form 4 buffer circuits). The pins that have the function of converting analogue signals to digital signals on the Lolin32 microcontroller There are also two transmission pins responsible for connecting to the OLED screen such as 21 and 22. The powers for sensor inputs, buffer, and display circuit are all taken from the 3.3 V source on the microcontroller. Signals from the pressure sensors go through the pressure divider and buffer circuits to the microcontroller, the communication process between the microcontroller and the computer is performed by the Arduino IDE software. In addition, data is also saved into CSV (Comma-separated values) files in Python for ease of use.

### *2.1.3 Data Collection*

Five healthy volunteers (four men – Aged: 27, Height [cm]: $170 \pm 5$, Weight [kg]: $65 \pm 8$, and one woman - Aged: 27, Height [cm]: 157, Weight [kg]: 50) participated in the experiment. More than, all volunteers were in normal physical condition and had no lower back pain previously. Eight postures collected include Empty seat, Upright sitting, Slouching, Leaning left, Leaning right, Left leg crossed, Right leg crossed, and Leaning backward as Figure 4.

Volunteers will sit in the order of postures described above, data collection time for each posture is one minute. After each position, the volunteer will stand up, at this time the microcontroller will collect signals from an empty chair, helping distinguish positions more easily.

To consider whether the placement of sensors is plausible in the classification of sitting posture, we observed the pressure distribution of ten sensors for seven postures in the early stage. Images of sitting postures and pressure maps of a volunteer with a height of 170 cm and weight of 65 kg are shown in Figure 4.

It is clear that the set of pressure values for each posture is completely different, classification of postures is promising in return. To be more specific, for upright sitting, the data is distributed quite evenly, especially for sensors closer to a pelvic area like S4, S5, S6, and S7 pose similar values and higher than the others. The slouching state records an uneven distribution of pressure. Because the angle created by the lumbar spine and mattress surface is less than 90 degrees, the pressure figures of the outside sensors like S3, S4, S6, and S7 are smaller than the case of sitting straight. When sitting on the left side, gravity exerts more on this side, so the left sensors record higher force values than the right. Depending on the inclination of the volunteer, the values of the sensors located on the left and right will obtain different values together. Therefore, we try to collect data at many different tilts to ensure the feasibility of the device in practical situations. In contrast, in the right-leaning, the pressure is mainly exerted on sensors on the right side. In the case of the left leg crossed, because one part of the left thigh lifts off the mattress surface, data from S1 will come in 0, S2 also records a decrease in signal. Meanwhile, S4 and S5 records receive higher values than their numbers in the case of the right-leaning position, although these two positions have pressure concentrated on one side accordingly. The value of sensor number 10 in the case of crossing the right leg on the left one is also small. When the right leg is raised, the pressure on the pelvic area of that leg increases, so the sensors located near this area record an increase in value compared to when leaning right. When sitting backward, the buttocks and pelvis slide slightly forward, so sensors placed behind such as S3, S4, S7, and S8 will record lower values when compared to the upright position or arching your back.

*2.2 Heart rate measuring module*

The heart rate is gauged by the Photoplethysmography technique, a non-invasive method that is used to measure changes in blood volume in subcutaneous microvascular tissue as changes in the intensity of light and therefore measure the heart rate. In the circuit design, we use the reflective principle to limit noise caused by motion and refer to the research of Aminah Hina et al. [54]. The principle diagram of the module is illustrated in Figure 5 and the ECG waveforms after each processing step are displayed in Figure 6, accordingly. The module includes a near-infrared LED, a photodiode and a processing unit. When it is worn on the user's wrist, light from the LED of this module is reflected by microvascular tissue and is collected by the photodiode. This photodiode generates a current proportional to the fluctuations in light intensity. The brightness of the LED is kept constant so that changes in light intensity are only caused by blood flow in the microvascular tissue. The current from the photodiode is fed to a trans-impedance amplifier (TIA) to convert to voltage and its form is shown in Figure 6 – (a). It is clear that this raw signal has a small amplitude and is disturbed by DC and high-frequency components. Therefore, it is passed into direct current removal (DCR) to remove the DC component caused by the respiratory signal and ambient light. Then, a non-inverting amplifier circuit (AMP) amplifies the signal. As a result, the ECG waveform is displayed in Figure 6 – (b). It is apparent that the signal is more smooth than raw data but its amplitude is still small. Then, this signal is filtered by the bandpass and its waveform is illustrated in Figure 6 – (c). The bandpass filter helps the signal to be clearer, showing its cyclic repetition which shows the heart rate at a degree. Because this work aims to use the Lolin32 microcontroller to display the heart rate and the sitting postures simultaneously, signals like those in Figure 6 – (c) cannot be satisfied with reading and finding the location of wave peaks. Therefore, we continue to use an inverting amplifier along with the circuit of a low-frequency filter with a cutoff frequency of 15 Hz to amplify amplitude and remove potentially harmful high-frequency component interference. The last step is further amplification by a programmable gain amplifier (PGA) and a clear ECG waveform appears in Figure 6 – (d). With the advantages of this signal, heart rate is easily calculated and converted into digital ones. IC LM324 is used in the filter and amplifier circuits. The signal is fed into the Lolin32 microcontroller to read and display real-time heart rate values.

### *2.3 Data processing*

Collecting data in alternating order of postures and empty seat status helps us save a lot of time in labelling and processing data. A total of 7205 samples are recorded including postures and empty chair data. The classification consists of ten features, so it cannot be displayed in 2- or 3-dimensional space properly. In the early step, we use two efficient methods, PCA (Principal Component Analysis) and t-SNE (t- Distributed Stochastic Neighbor Embedding) to visualize pressure distribution data while sitting.

#### *2.3.1 Training Models*

This work used four machine learning models; ANN (Artificial Neural Networks), SVM (Support Vector Machine), DT (Decision Tree), and RF (Random Forest); to train and evaluate the data with 80% of the data used in the training, the remaining 20% is for test process, and one-hot encode.

#### *2.3.2 Model Evaluation*

Once the model has been built and trained, evaluation of the test set is critical to test models. Evaluation criteria in this work include accuracy, confusion matrix, and $F1$-score indexes.

*Accuracy:* is used to evaluate by correctly predicted samples compared to total samples:

$$accuracy = \frac{TP + TN}{P + N}$$

Where TP is True Positive, TN is True Negative, P is Positive, and N is Negative.

*Confusion matrix:* is a table illustrating the performance of the model. It provides information about True and Predicted classes as follows:

| True Class<br>Predicted Class | Positive | Negative |
|---|---|---|
| **Positive** | **TP** | **FP** |
| **Negative** | **FN** | **TN** |

F1-Score is defined by Eq.

$$F_1 = \frac{TP}{TP + \frac{FN + FP}{2}}$$

FP is False Positive, FN is False Negative.

*2.4 Real-time module*

The selected model after training will be embedded into the Lolin32 microcontroller through the Micromlgen library. The results of heart rate and posture classification are displayed on an OLED screen by an I2C. Along with displaying on the OLED screen, the results are also shown on the mobile phone via Bluetooth. By connecting Bluetooth between the device and phone, users can check and monitor the posture and heart rate of the person sitting on the chair remotely and promptly remind them to sit correctly. Moreover, data can also be saved for analysis, statistics, and reporting to the users about their sitting postures and heart rate. From that, they adjust their sitting habits correctly in the future.

## 3. RESULTS

*3.1 Data visualization*

Data visualization plays an important role in providing a general view of data in terms of data structure, trends, and outliers. Then, the implementer can provide an appropriate layout for data processing based on it. As mentioned in the above part, the experimental data is visualised by two methods, PCA and t-SNE.

According to Figure 7, we easily see that sitting postures have little overlap, both 2-dimensional and 3-dimensional. Therefore, the accuracy of the classification model will be predicted to be high, and performance time will be short. Besides, the t-SNE has a longer processing time than PCA, but its results are more intuitive in both 2-dimensional and 3-dimensional space.

*3.2 Classification results*

There are 5764 samples for the training process and 1441 others for the testing process. Four different classification models were applied to find the model that has the highest accuracy. The main criteria used to evaluate the performance of models are accuracy and confusion matrix. Besides, $F_1$ index is used to compare models having the same accuracy.

The Artificial Neural Network (ANN) model with 10 input data layers and 8 output result layers uses the ReLu activation function for fast convergence, and to optimise calculation time. The model's accuracy reached 99.13%, and the confusion matrix is depicted in Figure 8 – (a). The confusion matrix also shows some samples incorrectly classified, for example, three postures of the right leg crossed over the left leg are mistakenly identified as slouching. However, these false samples are very few, and the model's accuracy is still over 99%.

With ten features classified in this case, Support Vector Machine (SVM) with the soft margin, kernel functions “Linear” along with the hyperparameter C=1 is used. The model provides an accuracy of 99.3% and avoids overfitting in classification. The confusion matrix of the model is displayed in Figure 8 – (b). As we can see, the accuracy obtained from the SVM model is higher than the artificial neural network (ANN) model. The mispredicted samples are also scattered. The index $F1$ of the model is 0.993.

The decision tree (DT) model illustrates an accuracy of 99.51%, and index $F1$ equals 0.9951, higher than the SVM. Its confusion matrix is shown in Figure 8 – (c).

The accuracy of the Random Forest model is 99.58%, the highest among the four models. Its confusion matrix also has better performance as Figure 8 – (d).

From the classification results of the four above machine learning models, there is no doubt that the ANN model has lower accuracy than traditional classification models. This is predictable because ANN needs a lot of data to achieve high performance. The deeper the network is, the larger the data is. With smaller data, traditional models outperform ANNs. In addition, traditional models do not require too many resources and computing power of the microcontroller to train the model. Therefore, the Random Forest model that gives the highest accuracy is chosen to be put into the Lolin32 microcontroller to detect the postures in real-time.

*3.3 ECG results*

To test the accuracy of our ECG measurement system, this work also chose a commercial device called DFRobot for comparison [55]. Recording the heart rate of the same volunteer by wearing them on two wrists, and their results are shown in Figure 9.

As a closer look at Figure 9, the heart rate values from our PPG module and DFRobot's sensor are highly correlated with the correlation coefficient of 0.965. To further evaluate the similarity of results between the measurement methods, the Bland-Altman is plotted in Figure 10.

According to Figure 10, the difference in heart rate value between the two measurement methods is 3 beats/minute and most of the different values are within the approximate limit. Hence, our proposed system demonstrated the feasibility and practicality of implementing a real-life application.

As illustrated above, the codes of the RF model and the heart rate calculation program are input into the Lolin32 microcontroller to display them in real-time. Then, the microcontroller is connected to external devices like mobile phones and laptops through the Serial Bluetooth Terminal and USB cable, respectively. This will help users read their postures and heart rate on the screen. Figure 11 is the real images of the sitting postures and the heart rate on the mobile phone and the laptop.

One utility this device brings to users is storing data about the heart rate and sitting posture into display modules like mobile phones or laptops. Therefore, users can pause the system, and select the mode of analysis to review the sitting behaviours and the heart rate. If users want to do the statistics about time intervals for each sitting posture they sat in, they will use the statistic button to display it like Figure 12. This allows them to make appropriate adjustments for their next sitting session.

## 4. Discussions

A smart chair with an invisible sensing system to monitor users' sitting postures and heart rates is necessary for residents, especially younger adults and office workers. The proposed system in this work mainly covers enough criteria for practical implementation. The main advantage of this system is its flexibility. The separate cushion helps users bring it anywhere they want. This is done because this work only uses the pressure sensor. We only place pressure sensors on the seat cushion without other places. In the literature, some research said that only a pressure mat on the seat can capture minor changes due to a switch of sitting posture [30]. Table 1 depicts some studies using a pressure array and comparison with our work. According to Table 1, our proposed method illustrates a higher accuracy than most other works although the number of sensors in our work is the least. One noticeable result is the contribution of Zhe Fan et al. [49]. They applied the CNN model to classify data and an accuracy of 99,82% is published. The main reason for the high results is that the collecting system is high-cost with a 44x52 pressure sensor array. Moreover, the large scale of data helps the training model to reach high accuracy. However, they did not design their method in real-time. This demonstrates that it is not necessary to place numerous sensors on the seat to sense changes in pressure when changing sitting posture. A suitable number of sensors along with appropriate placement of sensors also brings effective classification. This brings many merits in practical situations such as low cost, rapid processing, portability and real-time application.

We also make a comparison with the system using a few sensors. From our observation, few sensor-based systems usually combine different types of sensors like infrared reflective distance sensors [55], load cells [37] and so on. Some results are shown in Table 2.

A closer look at Table 2 shows that our accuracy is much higher than most other research. According to [30, 58], one type of sensor used to classify the sitting posture does not bring good performance. Hence, mounting sensors on the seat and backrest is necessary to sense enough information about posture changes. However, their results are low while our results are higher. This can be explained that sensor distribution and the number of sensors in our work are optimal, enough to capture minor changes in sitting posture. In general, our design optimizes the number of sensors, leading to cost-effectiveness, easy processing, and computational time-effective. The separate cushion of this work also brings convenience to users. From that, it is a promising system to implement in practical situations.

Regarding the ECG module, Pereira L. et al. [39] showed an accuracy of 90.50% in peak catching. Po-Cheng Su et al. [52] used capacitive coupling of textiles to measure ECG signals with two sheets of the conductive textile being mounted on the back and the seat to record the heart data. They did not show their accuracy and only provided signal-noise ratio (SNR). In our work, we compare our results with the commercial sensor DFRobot and it is identical. This is an advantage for us to spread the system of smart cushion and the heart rate module.

## 5. Conclusions

Our work proposes an optimal invisible module to monitor sitting posture with numerous of advantages. Firstly, using a few sensors reduces production expenses, and eases computational processing. Secondly, the pressure-only-based system creates a portable cushion, therefore being advantageous for bringing. Thirdly, the accuracy is significantly high compared to most previous research. This is a promising signal for applying it in real life. Fourth, the heart rate module shows an identical result to the commercial sensor DFRobot, counting the heart rate accurately. Finally, the heart rate and sitting posture are displayed on external devices like mobile phones, laptops and so on with many associated convenient functions. Users can see information about their sitting postures and heart rate saved in external devices, and do the statistics to have general information about their sitting habits. From that, they can adjust them to improve their health in the long run. However, this work should supply some necessary duties like changing the number of sensors to identify the optimal quantity of sensors. Although the heart rate module is visible, utilities integrated into the processing unit such as reviewing the heart rate and doing the statistics to look at the relationship between the sitting postures and heart rate that this system brings help users have a general view of their habits. From that, they will adjust them to have better health.

## 6. Authors' contribution

Vo Quoc Bao: Software, Validation, Resources, Data Curation, Methodology. Nguyen Thi Minh Huong: Investigation, Writing- Original draft preparation, Writing - Review & Editing, Visualization. Huynh Quang Linh: Supervision. Nguyen Trung Hau: Conceptualization, Formal analysis, Project administration, Funding acquisition.

**Acknowledgements**

This research is funded by Vietnam National University HoChiMinh City (VNU-HCM) under grant number C2022-20-03.

## References

[1] S. Parry, M. Chow, F. Batchelor amd R. E. Fary, Physical activity and sedentary behaviour in a residential aged care facility. *Australasian journal on ageing* vol. 38, E12-E18, 1 (2019).

[2]  JL. Kelsey, An epidemiological study of the relationship between occupations and acute herniated lumbar intervertebral discs. International Journal of Epidemiology, vol. 4, no. 3, pp. 197–205 (1975).

[3] A. Magora, Investigation of the relation between low back pain and occupation. 3. Physical requirements: sitting, standing and weight lifting. Industrial medicine and surgery, vol. 41, no. 12, pp. 5 – 9 (1972).

[4] JW. Frymoyer and WL. Cats-Baril, An overview of the incidences and costs of low back pain. Orthopedic Clinics of North America, vol. 22, no. 2, pp. 263–271 (1991).

[5] H R Guo, S Tanaka, L L Cameron, P J Seligman, V J Behrens, J Ger, D K Wild and V Putz-Anderson, Back pain among workers in the United States: National Estimates and workers at high risk. American Journal of Industrial Medicine, vol. 28, no. 5, pp. 591–602 (1995).

[6] AC. Papageorgiou, PR. Croft, S. Ferry, ML. Jayson, and AJ. Silman, Estimating the prevalence of low back pain in the general population. Spine, vol. 20, no. 17, pp. 1889–1894 (1995).

[7] K. Walsh, M. Cruddas, and D. Coggon, Low back pain in eight areas of Britain. Journal of Epidemiology and Community Health, vol. 46, no. 3, pp. 227–230 (1992).

[8] K. O'Sullivan, M. O'Keeffe, L. O'Sullivan, P. O'Sullivan, and W. Dankaerts, Perceptions of sitting posture among members of the community, both with and without non-specific chronic low back pain. Manual Therapy, vol. 18, no. 6, pp. 551–556 (2013).

[9] K. Barczyk-Pawelec and T. Sipko, Active self-correction of spinal posture in pain-free women in response to the command 'Straighten your back'. Women Health, vol. 57, no. 9, pp. 1098–1114 (2016).

[10] AP. Claus, JA. Hides, GL. Moseley, and PW. Hodges, Is 'ideal' sitting posture real?: Measurement of spinal curves in four sitting postures. Manual Therapy, vol. 14, no. 4, pp. 404–408 (2009).

[11] P. Vermander, A. Mancisidor,  I. Cabanes and N. Perez, Intelligent systems for sitting posture monitoring and anomaly detection: an overview. J NeuroEngineering Rehabil 21, 28 (2024).

[12] A. Kulikajevas, R. Maskeliunas and R. Damaševičius, Detection of sitting posture using hierarchical image composition and deep learning. PeerJ. Comput. Sci., 7 (2021).

[13] X. Huang and L. Gao, Reconstructing Three-Dimensional Human Poses: A Combined Approach of Iterative Calculation on Skeleton Model and Conformal Geometric Algebra. Symmetry 11, no. 3: 301 (2019).

[14] J. Liu, Y. Wang, Y. Liu, S. Xian and C. Pan, 3D posturenet: a unified framework for skeleton-based posture recognition. Pattern Recognition Letters, 140(8):143–149 (2020).

[15] M. Tariq, H. Majeed, MO. Beg, FA. Khan and A. Derhab,  Accurate detection of sitting posture activities in a secure IoT based assisted living environment. Futur. Gener. Comput. Syst., 92, pp. 745-757 (2019).

[16] ES. Ho, JCP. Chan, DCK. Chan, HP. Shum, YM. Cheung and PC. Yuen, Improving posture classification accuracy for depth sensor-based human activity monitoring in smart environments. Comput. Vis. Image Underst., 148, pp. 97-110 (2016).

[17] T. Kong, W. Fang, PED. Love, H. Luo, S. Xu and H. Li, Computer vision and long short-term memory: learning to predict unsafe behaviour in construction. Adv. Eng. Informatics., 50, Article 101400 (2021).

[18] B. Liu, Y. Li, S. Zhang and X. Ye, Healthy human sitting posture estimation in RGB-D scenes using object context. Multimedia Tools Appl, 76(8):10721–39 (2017).

[19] Y. Zhang, Y. Huang, B. Lu, Y. Ma, J. Qiu, Y. Zhao, X. Guo, C. Liu, P. Liu and Y. Zhang, Real-time sitting behavior tracking and analysis for rectification of sitting habits by strain sensor-based flexible data bands. Meas. Sci. Technol., 31 (2020).

[20] L. Feng, Z. Li, C. Liu, X. Chen, X. Yin and D. Fang, SitR: sitting posture recognition using RF signals. IEEE Internet Things J., 7 (12), pp. 11492-11504 (2020).

[21] ND. Nath, T. Chaspari and AH. Behzadan, Automated ergonomic risk monitoring using body-mounted sensors and machine learning. Adv. Eng. Informatics., 38, pp. 514-526 (2018).

[22] J. Zhao and E. Obony, Convolutional long short-term memory model for recognizing construction workers' postures from wearable inertial measurement units. Adv. Eng. Informatics., 46, Article 101177 (2020).

[23] M. Zaltieri, DL. Presti, M. Bravi, MA. Caponero, S. Sterzi, E. Schena and C. Massaroni, Assessment of a multi-sensor FBG-based wearable system in sitting postures recognition and respiratory rate evaluation of office workers. IEEE Trans Biomed Eng;70(5):1673–82 (2023).

[24] F. Tlili, R. Haddad, R. Bouallegue and R. Shubair, Design and architecture of smart belt for real time posture monitoring. Internet of Things, 17 (2022).

[25] A. Cristina, F. Geraldo and AM. Kuasne, Prototype of wearable technology applied to the Monitoring of the vertebral column. Int J Online Biomed Eng, 16(01):34–50 (2020).

[26] S. Matuska, M. Paralic and R. Hudec, A smart system for sitting posture detection based on force sensors and mobile application. Mob. Inf. Syst. (2020).

[27] Q. Hu, X. Tang and W. Tang, A smart chair sitting posture recognition system using flex sensors and FPGA implemented artificial neural network. IEEE Sens. J., 20 (14), pp. 8007-8016 (2020).

[28] AR. Anwary, D. Cetinkaya, M. Vassallo and H. Bouchachia, Smart-cover: a real-time sitting posture monitoring system, sensors actuators. A Phys., 317 (2021).

[29] C. Ma, W. Li, R. Gravina and G. Fortino, Posture detection based on smart cushion for wheelchair users. Sensors., 17, pp. 6-18 (2017).

[30] J. Meyer, B. Arnrich, J. Schumm and G. Troster, Design and modeling of a textile pressure sensor for sitting posture classification. IEEE Sens. J., 10 (8), pp. 1391-1398 (2010).

[31] M. Kim, H. Kim, J. Park, KK. Jee, JA. Lim and MC. Park, Real-Time Sitting Posture Correction System Based on Highly Durable and Washable Electronic Textile Pressure Sensors. Sens. Actuators A Phys, 269, 394–400 (2018).

[32] W. Xu, MC. Huang, N. Amini, L. He and M. Sarrafzadeh, eCushion: A Textile Pressure Sensor Array Design and Calibration for Sitting Posture Analysis. IEEE Sens. J., 13, 3926–3934 (2013).

[33] M. Martínez-Estrada, T. Vuohijoki, A. Poberznik, A. Shaikh, J. Virkki, I. Gil and RA. Fernández-García, Smart Chair to Monitor Sitting Posture by Capacitive Textile Sensors. Materials, 16, 48382 (2023).

[34] F. Luna-Perejón, JM. Montes-Sánchez, L. Durán-López, A. Vazquez-Baeza, I. Beasley-Bohórquez and JL. Sevillano-Ramos, IoT Device for Sitting Posture Classification Using Artificial Neural Networks. Electronics, 10, 1825 (2021).

[35] C. Ma, W. Li, R. Gravina, J. Cao, Q. Li and G. Fortino, Activity level assessment using a smart cushion for people with a sedentary lifestyle. Sensors, 17 (10): 2269 (2017).

[36] S. Sifuentes, R. Gonzalez-Landaeta, J. Cota-Ruiz and F. Reverter, Seat occupancy detection based on a low-power microcontroller and a single FSR. Sensors, 19 (3): 699 (2019).

[37] C. Tavares, J. Silva, A. Mendes, L. Rebolo, DM. Fatima, N. Alberto, M. Lima, Radwan, HP. Da Silva and P. Antunes, Smart office chair for working conditions optimization. IEEE Access. (April):50497–50509 (2023).

[38] J. Roh, H. Park, K. Lee, J. Hyeong, S. Kim and B. Lee, Sitting Posture Monitoring System Based on a Low-Cost Load Cell Using Machine Learning. Sensors, 18, 208 (2018).

[39] L. Pereira and HA. Plácido Da Silva, Novel Smart Chair System for Posture Classification and Invisible ECG Monitoring. Sensors, 23, 719 (2023).

[40] Y. Otoda, T. Mizumoto, Y. Arakawa, C. Nakajima, M. Kohana, M. Uenishi and K. Yasumoto, Census: Continuous posture sensing chair for office workers. In Proceedings of the 2018 IEEE International Conference on Consumer Electronics (ICCE), Las Vegas, NV, USA, 12–14 January 2018.

[41] S. Ma, WH. Cho, CH. Quan and S. Lee, A sitting posture recognition system based on 3-axis accelerometer. Proceedings of the 2016 IEEE Conference on Computational Intelligence in Bioinformatics and Computational Biology (CIBCB); Chiang Mai, Thailand. 5–7; pp. 1–3 (2016).

[42] DF. Odesola, J. Kulon, S. Verghese, A. Partlow and C. Gibson, Smart Sensing Chairs for Sitting Posture Detection, Classification, and Monitoring: A Comprehensive Review. Sensors, 24(9):2940 (2024).

[43] M. Huang, I. Gibson and R. Yang, Smart Chair for Monitoring of Sitting Behavior. KEG, 2, 274 (2017).

[44] Q. Wan, H. Zhao, J. Li and P. Xu, Hip Positioning and Sitting Posture Recognition Based on Human Sitting Pressure Image. Sensors, 21, 426 (2021).

[45] Y. Kim, Y. Son, W. Kim, B. Jin and M. Yun, Classification of Children's Sitting Postures Using Machine Learning Algorithms. Appl. Sci., 8, 1280 (2018).

[46] X. Ran, C. Wang, Y. Xiao, X. Gao, Z. Zhu and B. Chen, A Portable Sitting Posture Monitoring System Based on a Pressure Sensor Array and Machine Learning. Sens. Actuators A Phys., 331, 112900 (2021).

[47] J. Ahmad, J. Sidén and H. Andersson, A Proposal of Implementation of Sitting Posture Monitoring System for Wheelchair Utilizing Machine Learning Methods. Sensors, 21, 6349 (2021).

[48] J. Wang, B. Hafidh, H. Dong and El Saddik, Sitting Posture Recognition Using a Spiking Neural Network. IEEE Sens. J., 21, 1779–1786 (2021).

[49] Z. Fan, X. Hu, W.-M Chen, D.-W Zhang and X. Ma, A Deep Learning Based 2-Dimensional Hip Pressure Signals Analysis Method for Sitting Posture Recognition. Biomed. Signal Process. Control, 73, 103432 (2022).

[50] Apple Watch, https://www.apple.com/watch/

[51] Dong ho thong minh, https://www.thegioididong.com/dong-ho-thong-minh-apple

[52] S. Po-Cheng, H. Ya-Hsin, K. Ming-Ta, C. Jyun-Jhe and L. Ping-Chen, Noncontact ECG Monitoring by Capacitive Coupling of Textiles in a Chair. Journal of healthcare engineering vol. 2021 6698567. 16 (2021).

[53] DE. Lee, SM. Seo, HS. Woo, and SY. Won, Analysis of body imbalance in various writing sitting postures using sitting pressure measurement. Journal of Physical Therapy Science, vol. 30, no. 2, pp. 343–346 (2018).

[54] A. Hina, H. Nadeem, and W. Saadeh, A single led photoplethysmography-based noninvasive glucose monitoring prototype system. 2019 IEEE International Symposium on Circuits and Systems (ISCAS), (2019).

[55] DFRobot, https://www.dfrobot.com/.

[56] H. Jeong, and W. Park, Developing and Evaluating a Mixed Sensor Smart Chair System for Real-Time Posture Classification: Combining Pressure and Distance Sensors. in IEEE Journal of Biomedical and Health Informatics, vol. 25, no. 5, pp. 1805-1813 (2021).

[57] T. Aminosharieh Najafi, A. Abramo, K. Kyamakya and A. Affanni, Development of a Smart Chair Sensors System and Classification of Sitting Postures with Deep Learning Algorithms. Sensors, 22, 5585 (2022).

[58] H. Tan, L. Slivovsky and A. Pentland, A sensing chair using pressure distribution sensors. Mechatronics, IEEE/ASME Trans, 6, 261–268 (2001).

[59] W. Cai, D. Zhao, M. Zhang, Y. Xu and Z. Li, Improved Self-Organizing Map-Based Unsupervised Learning Algorithm for Sitting Posture Recognition System. Sensors, 21, 6246 (2021).

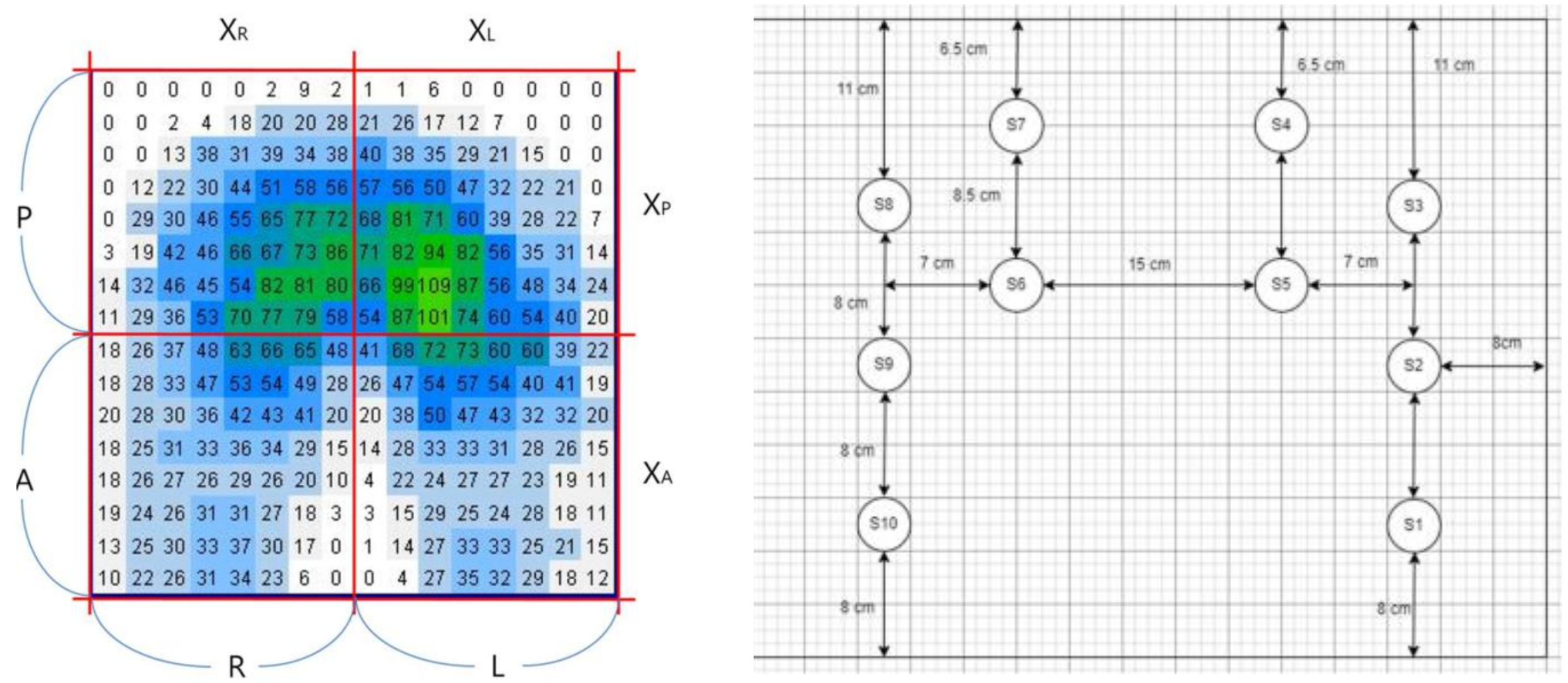

(a) Pressure mapping according to [53]

(b) Placement of sensors on the surface of the cushion

Figure 1. The pressure mapping and sensor distribution

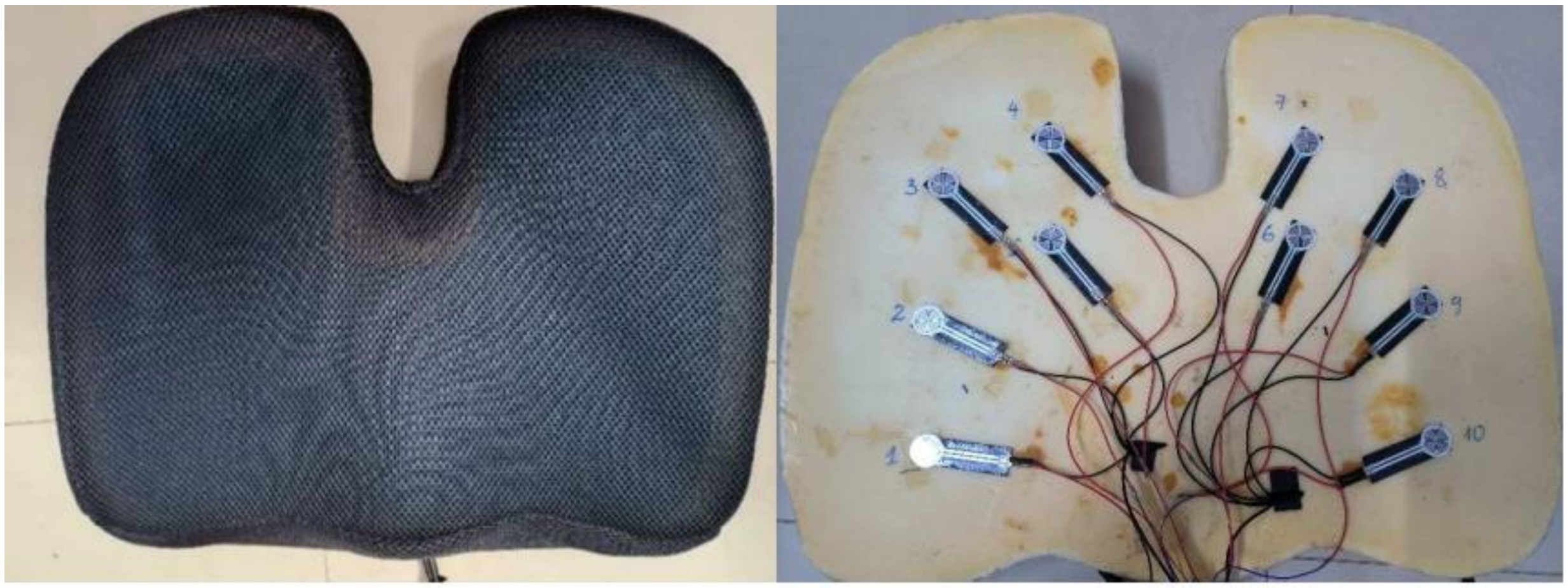

Figure 2. A real image of sensors on the mattress covered by the wrapper of the seat cushion

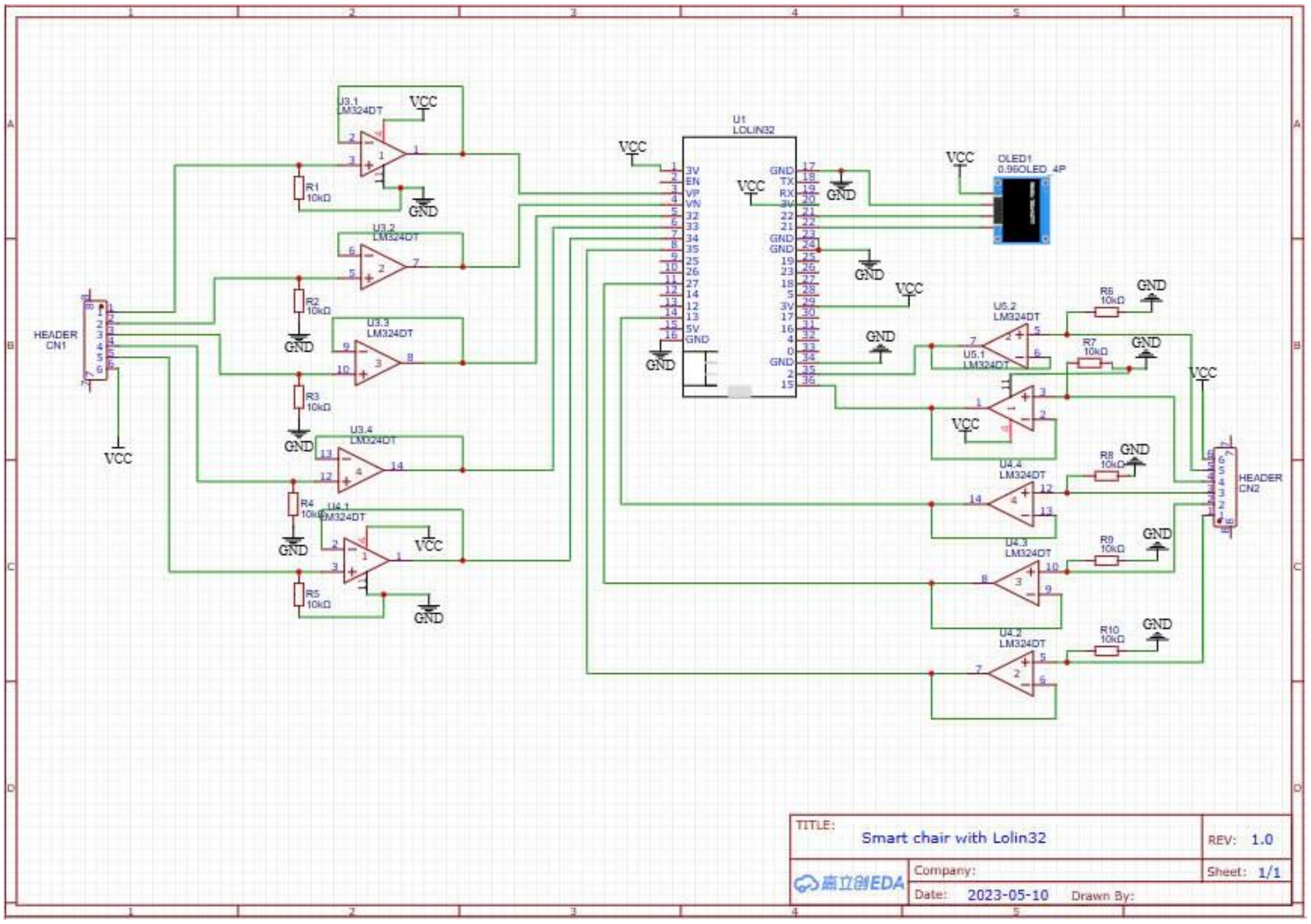

Figure 3. The principle diagram of the sitting posture classification module

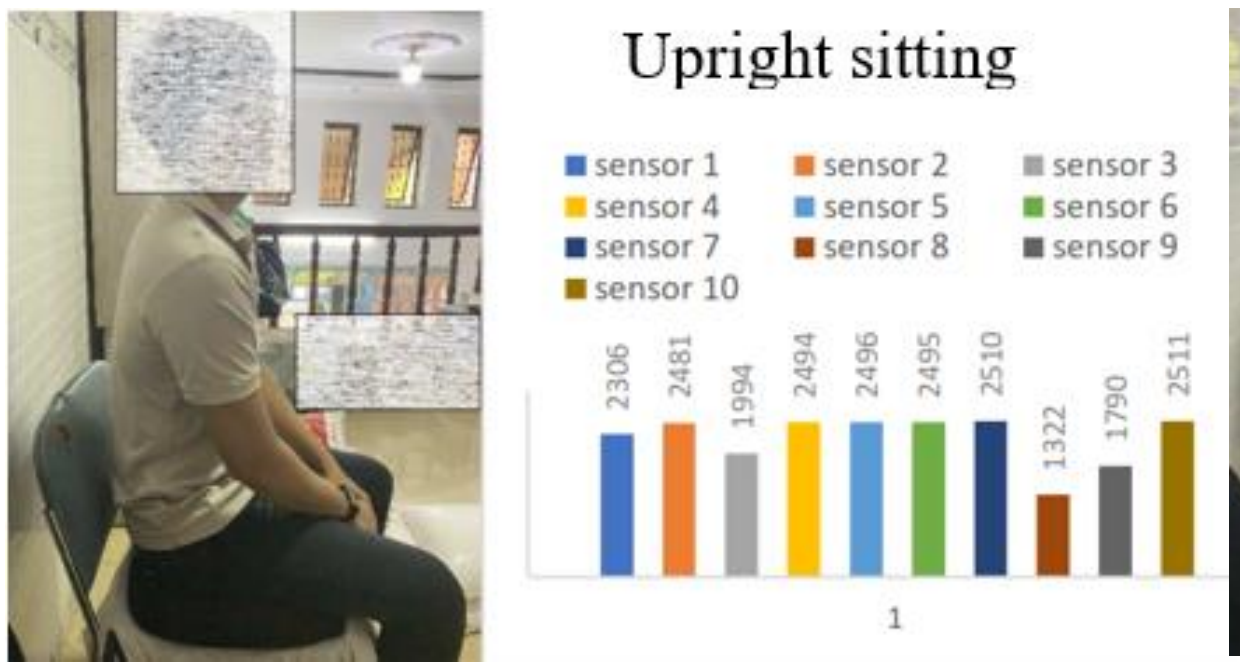

(a) Upright sitting and its pressure map

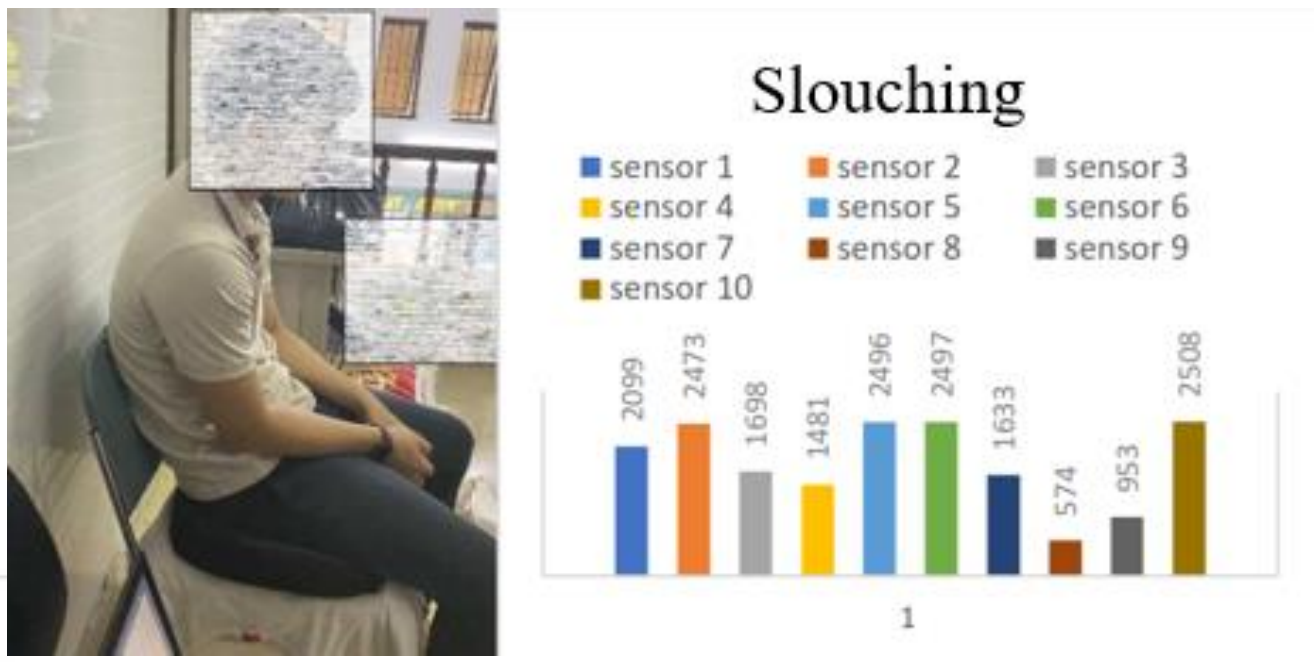

(b) Slouching and its pressure map

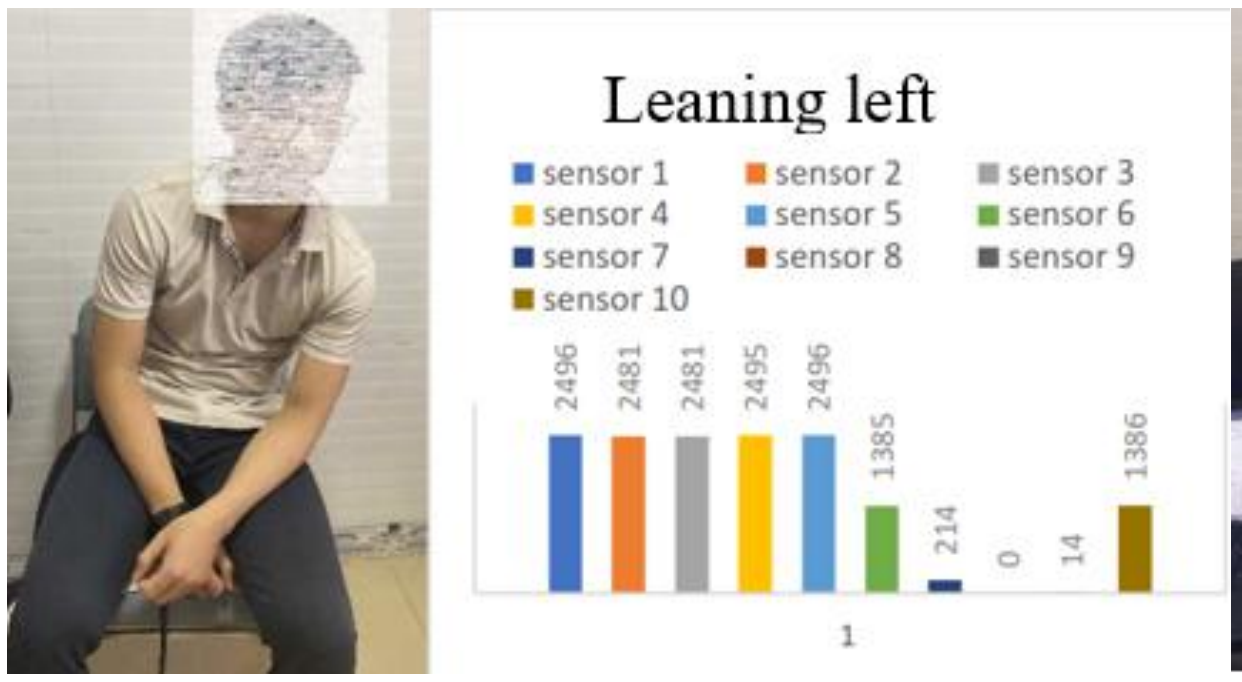

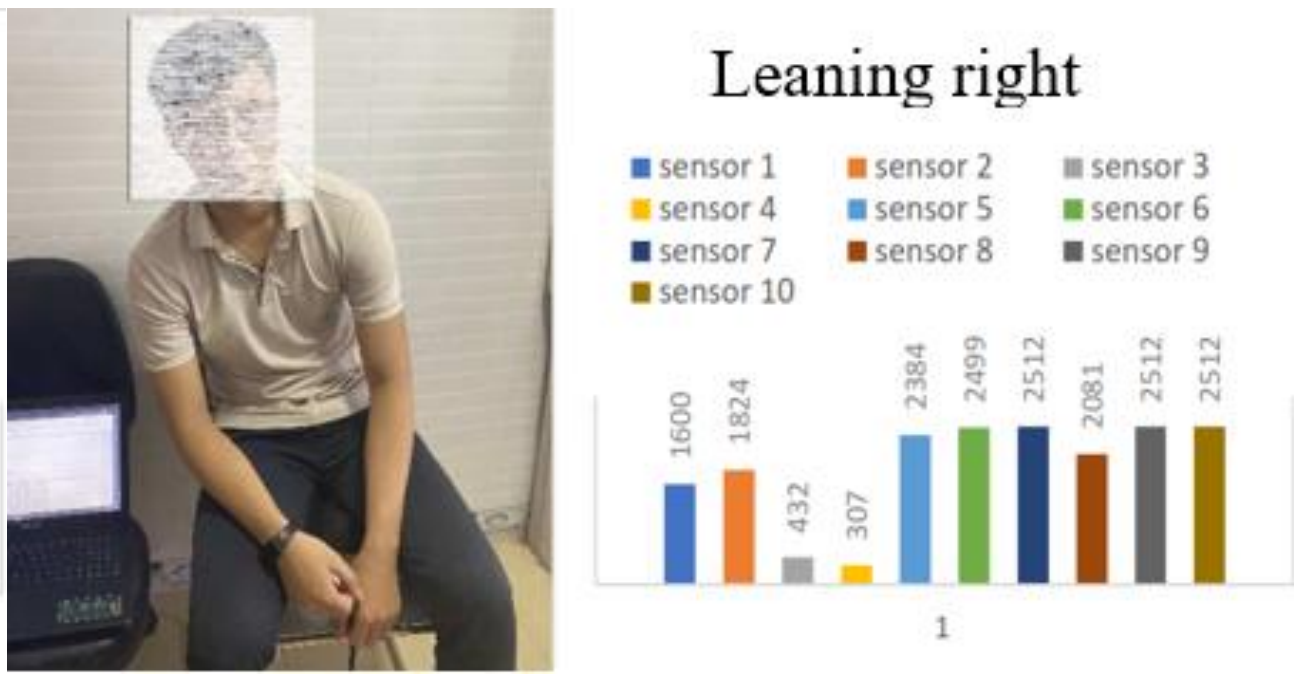

(c) Leaning left and its pressure map

(d) Leaning right and its pressure map

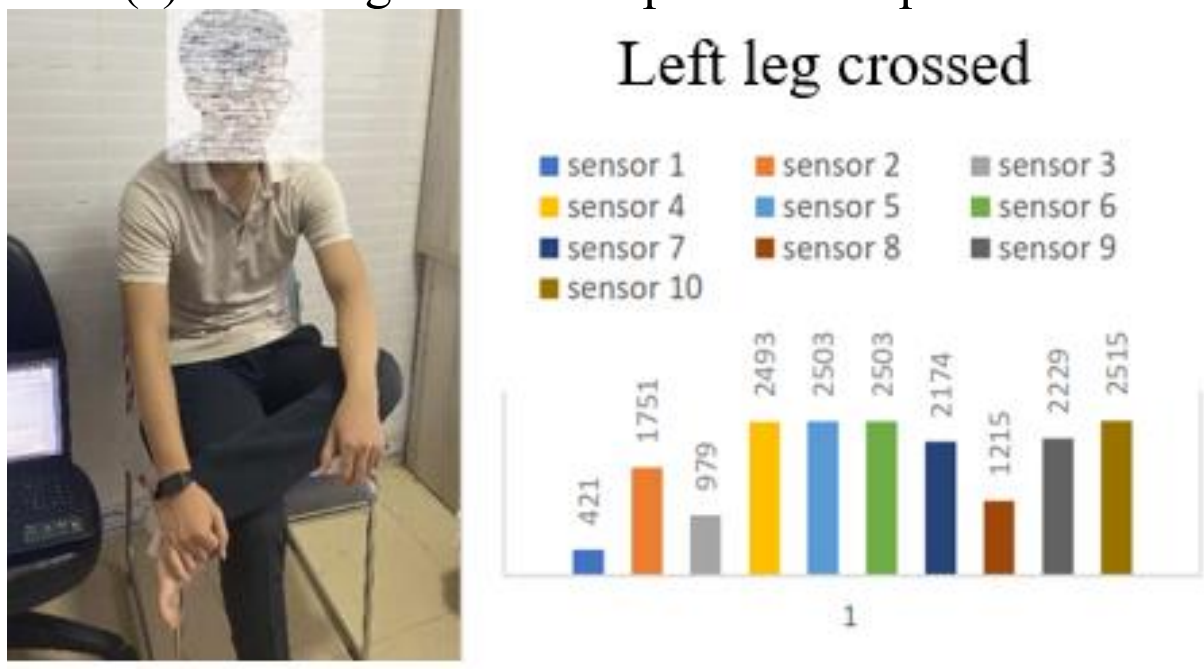

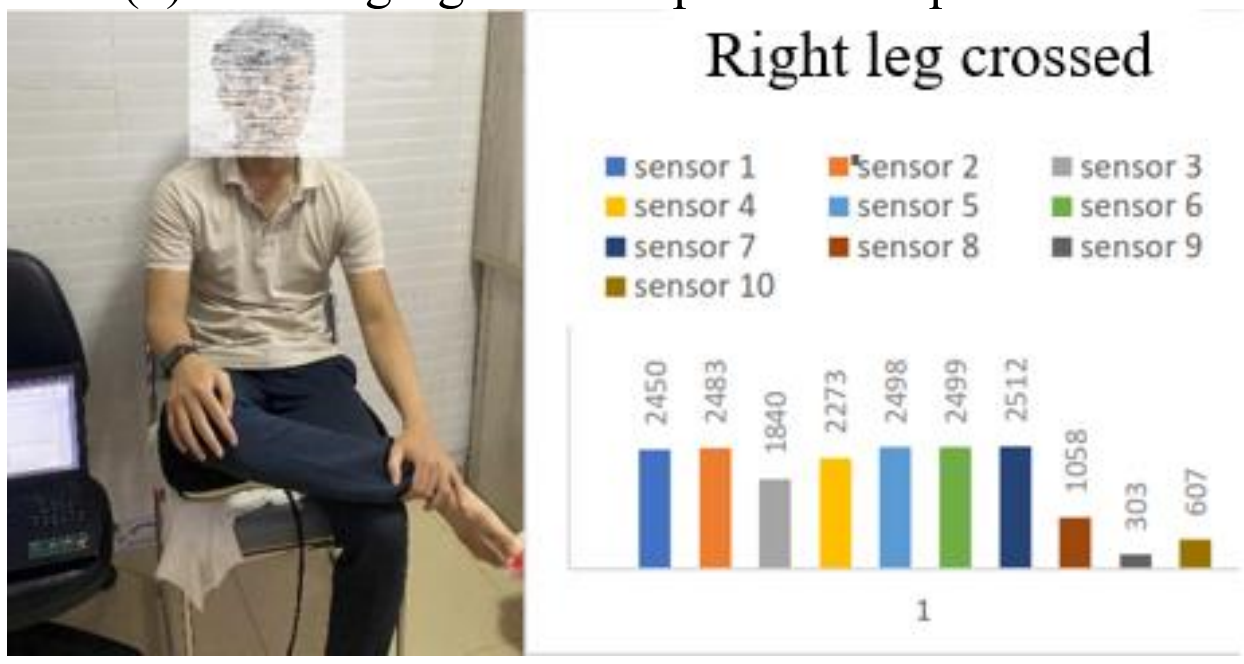

(e) Left leg crossed and its pressure map

(f) Right leg crossed and its pressure map

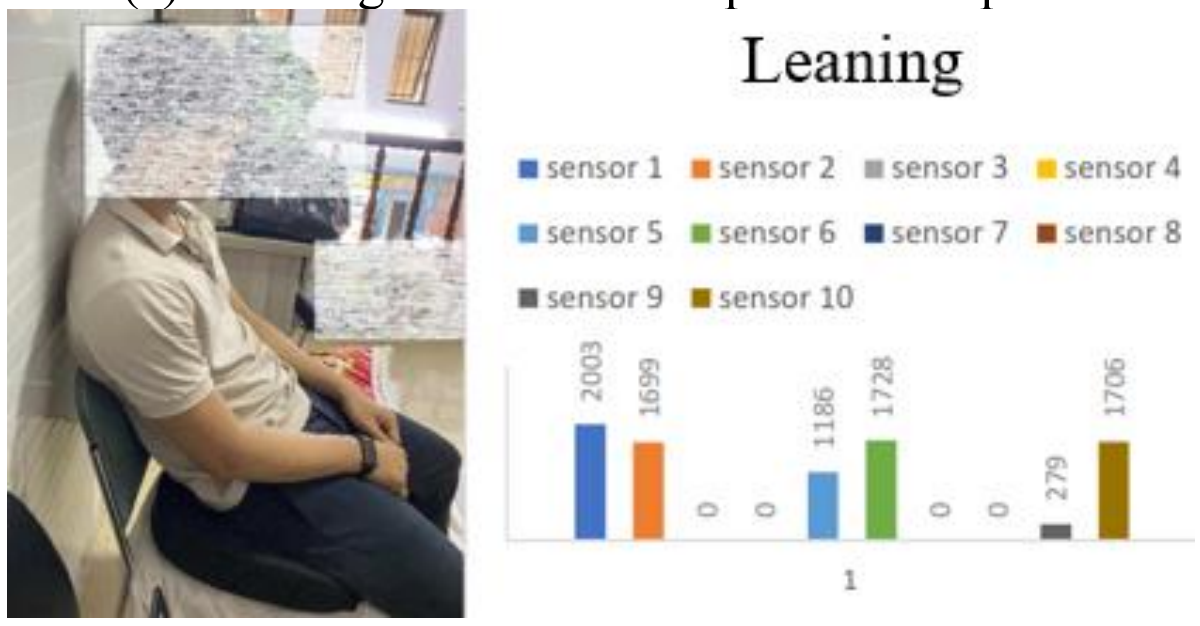

(g) Leaning and its pressure map

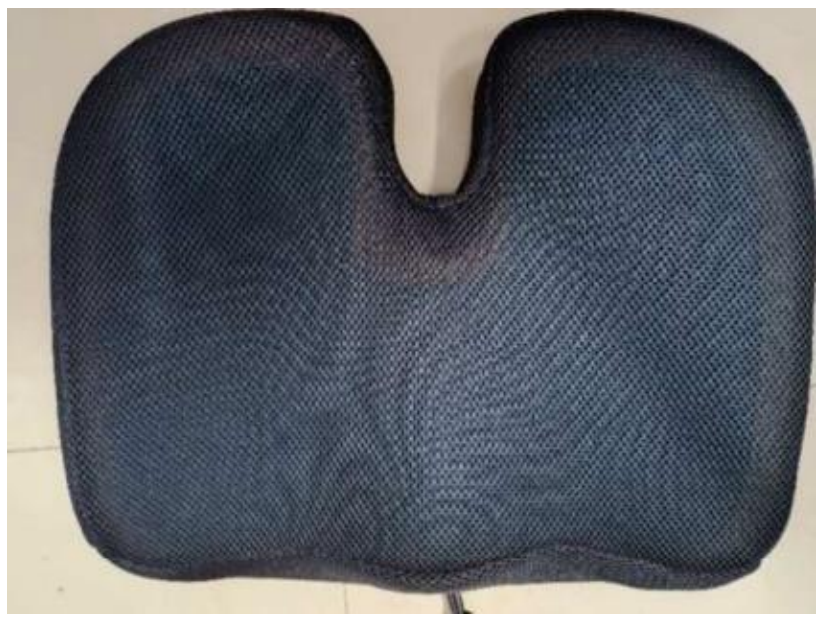

(h) Empty seat

Figure 4. The flux diagram of the collection process

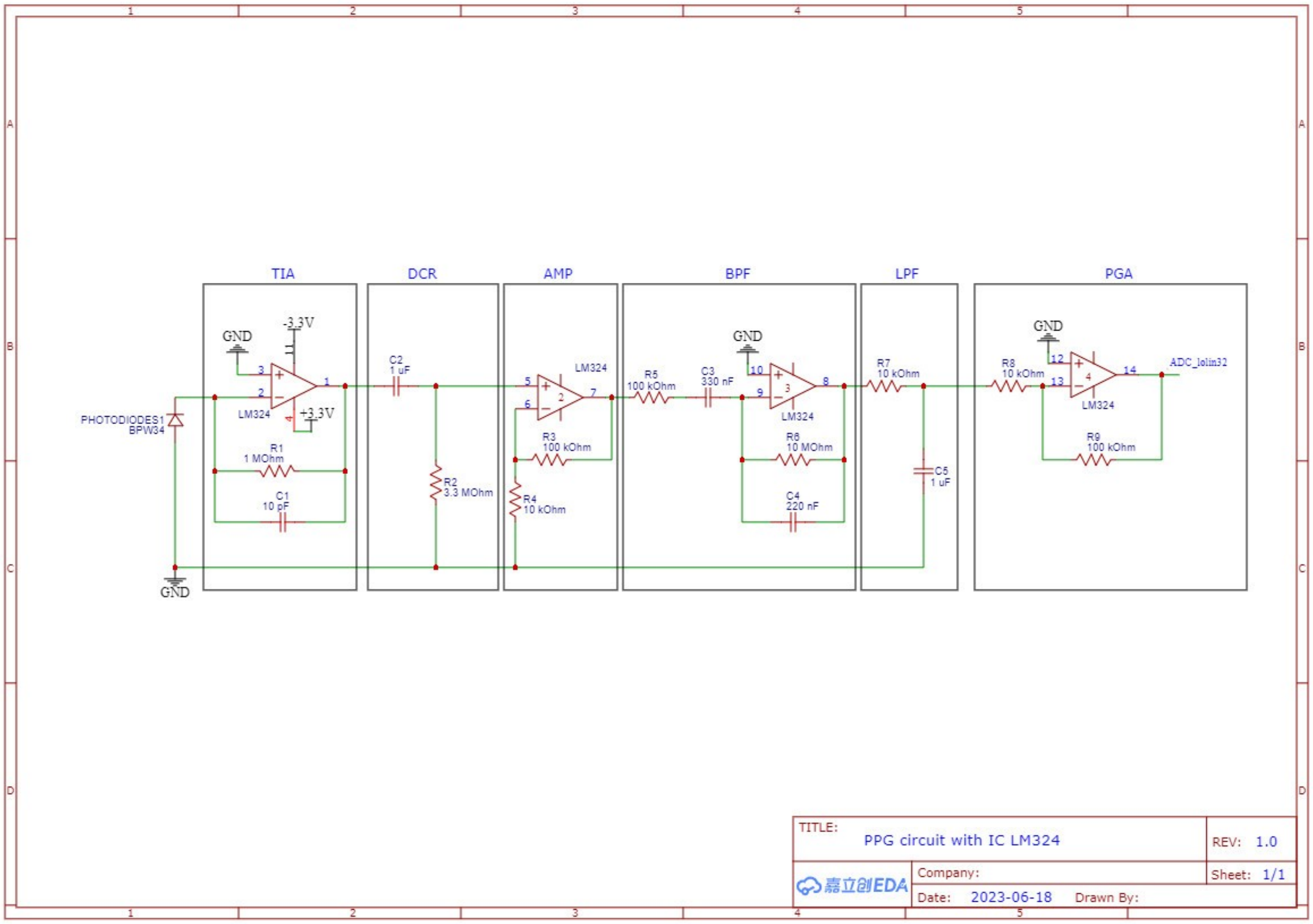

Figure 5. The schematics of the PPG circuit.

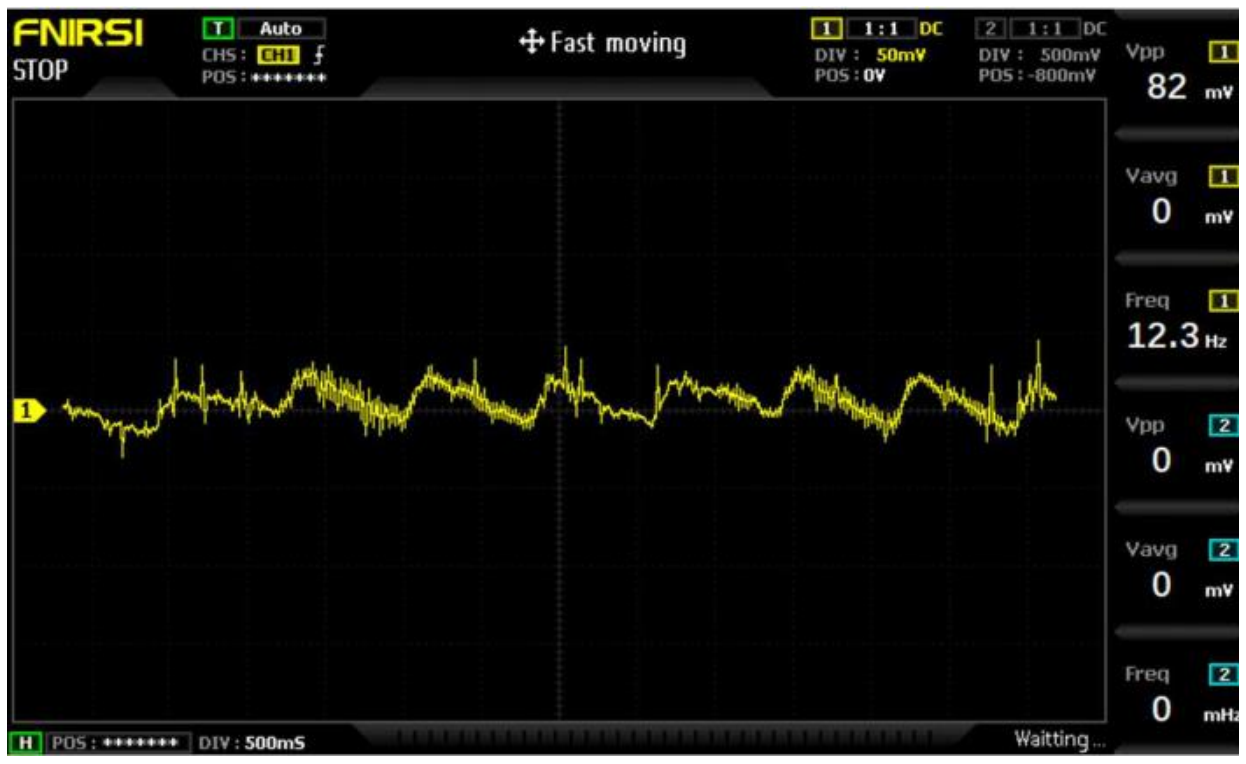

(a) Waveform of ECG after passing through the TIA circuit

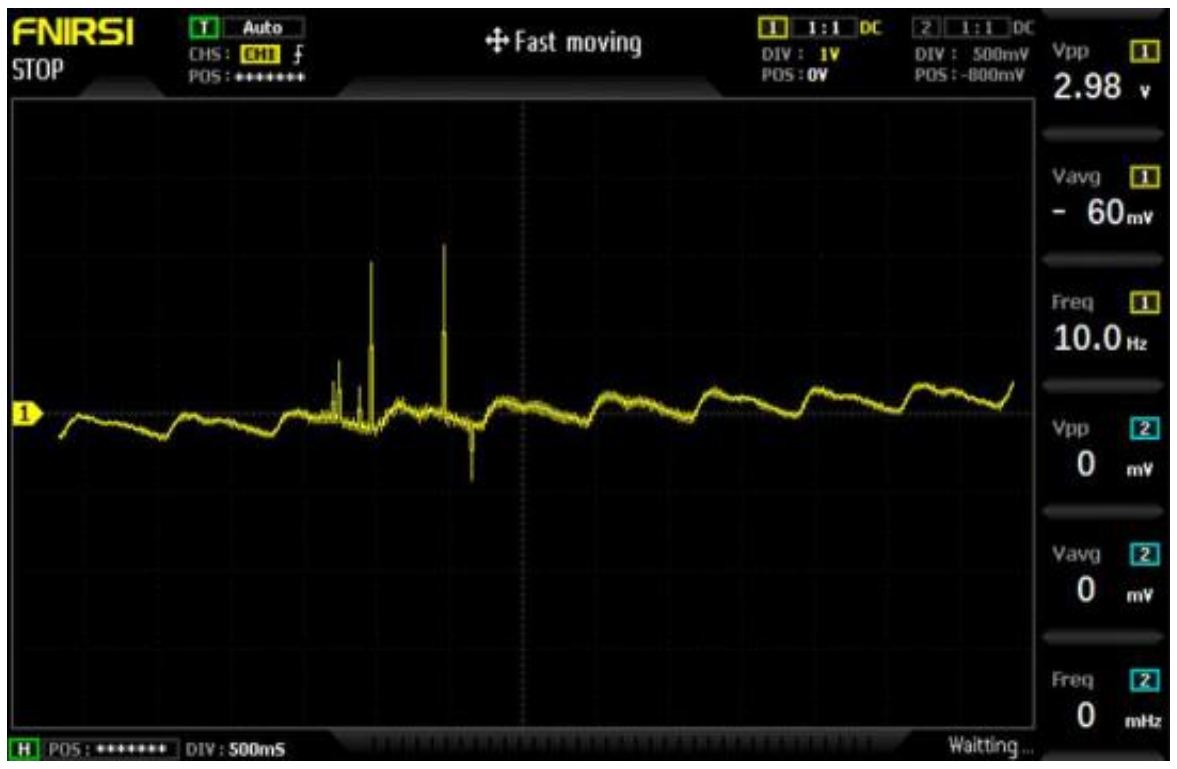

(b) Waveform of ECG after removing DC component and is firstly amplified

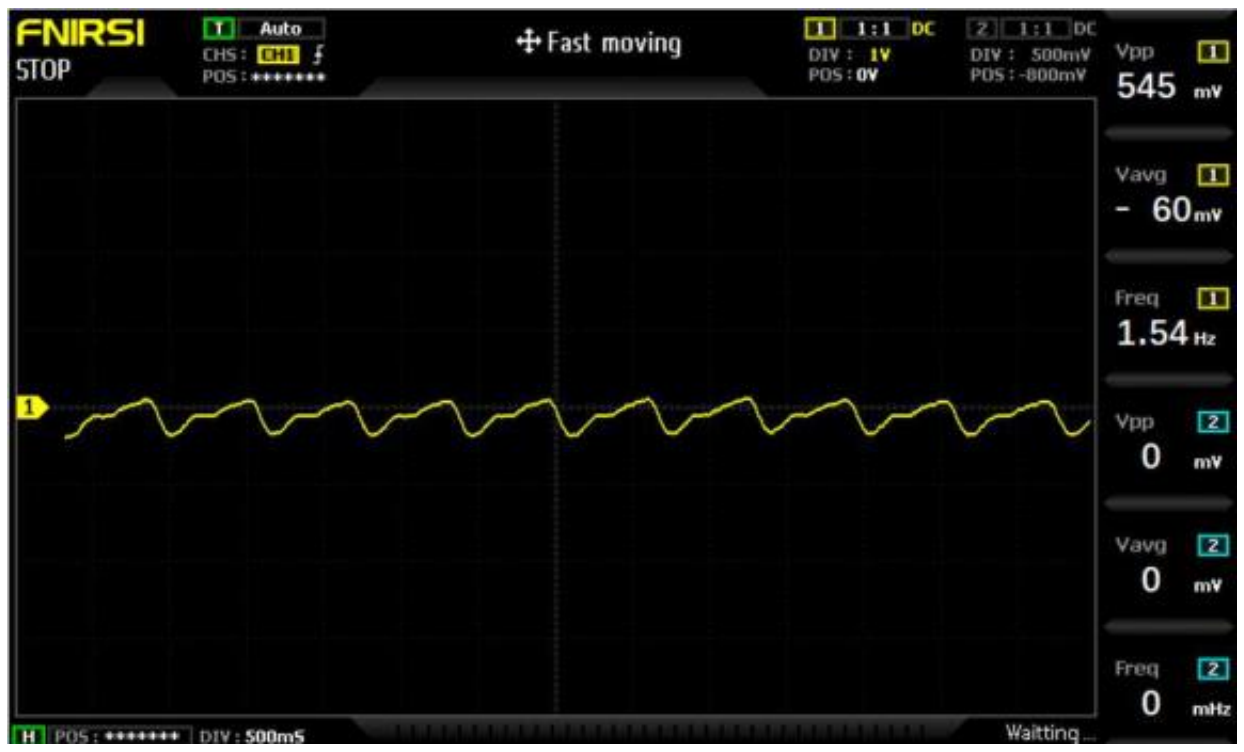

(c) ECG signal after filtering by the bandpass

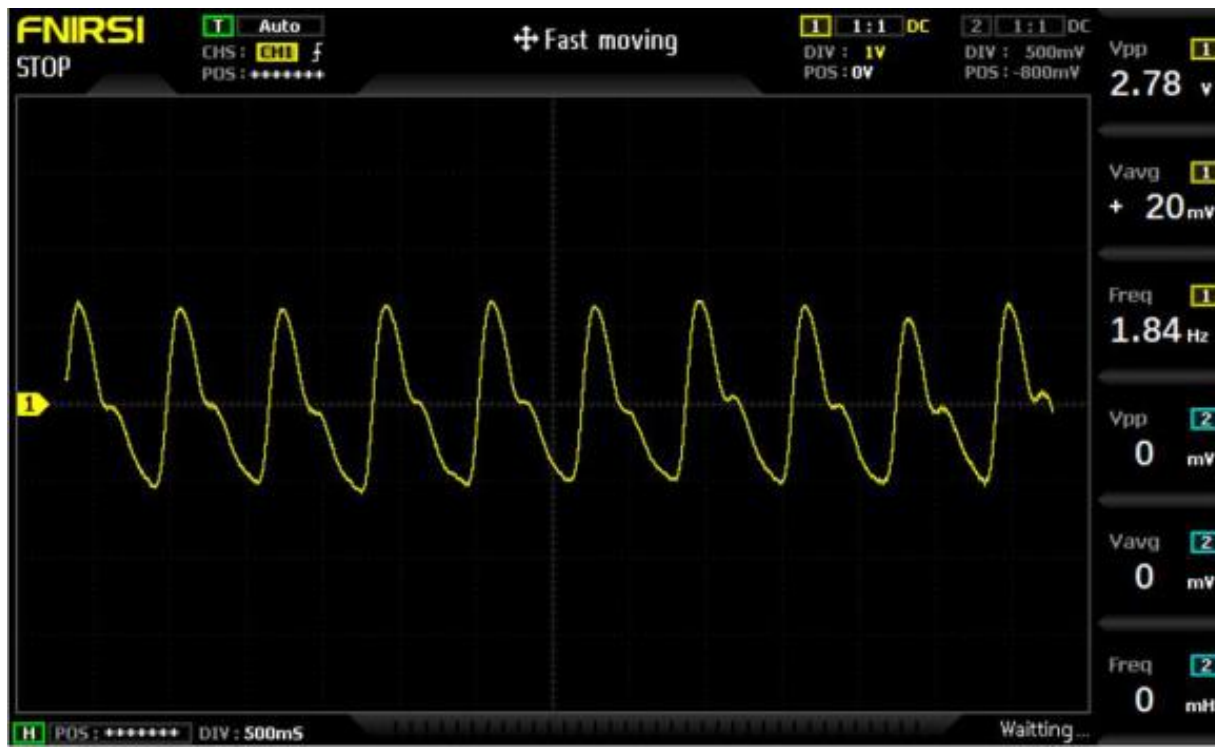

(d) ECG signal before microcontroller.

Figure 6. Waveform of ECG after each processing unit of the PPG circuit.

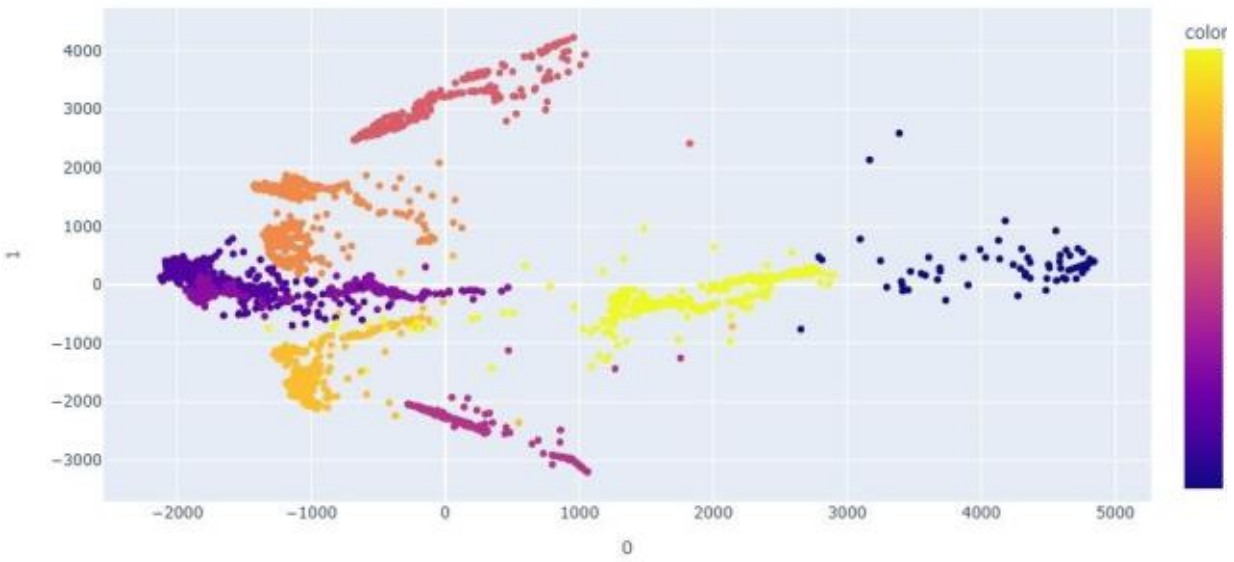

(a) Data visualization by PCA 2D

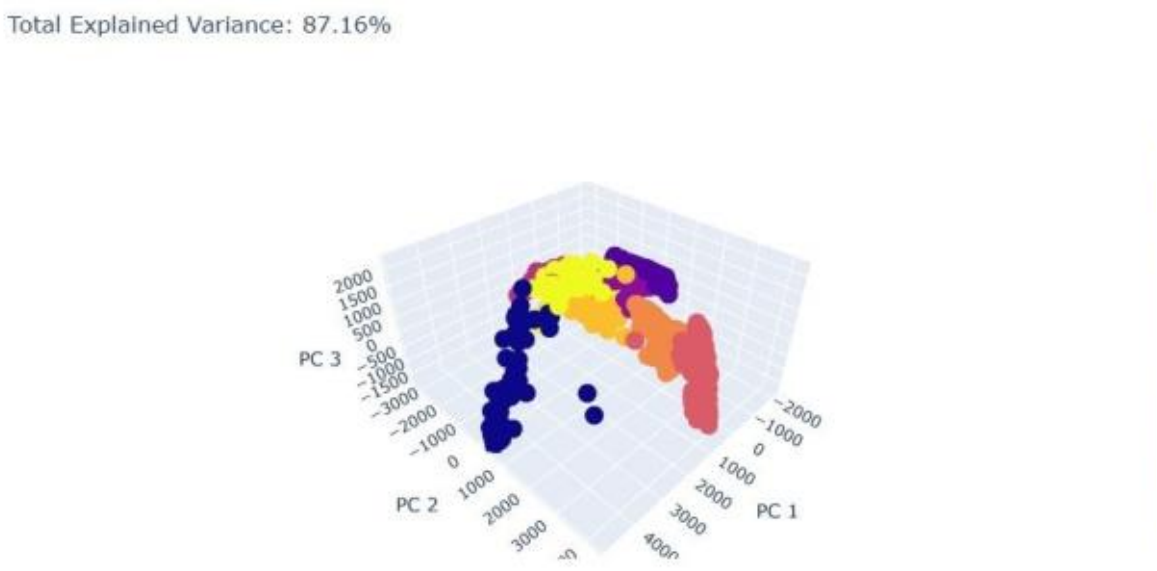

(b) Data visualization by PCA 3D

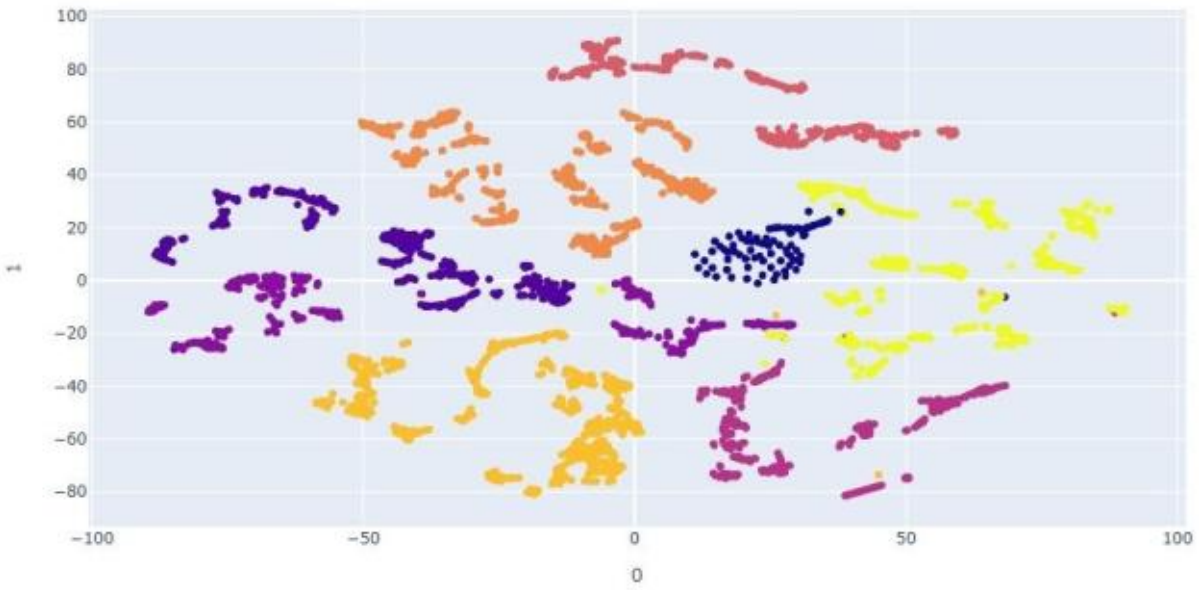

(c) Data visualization by t-SNE 2D

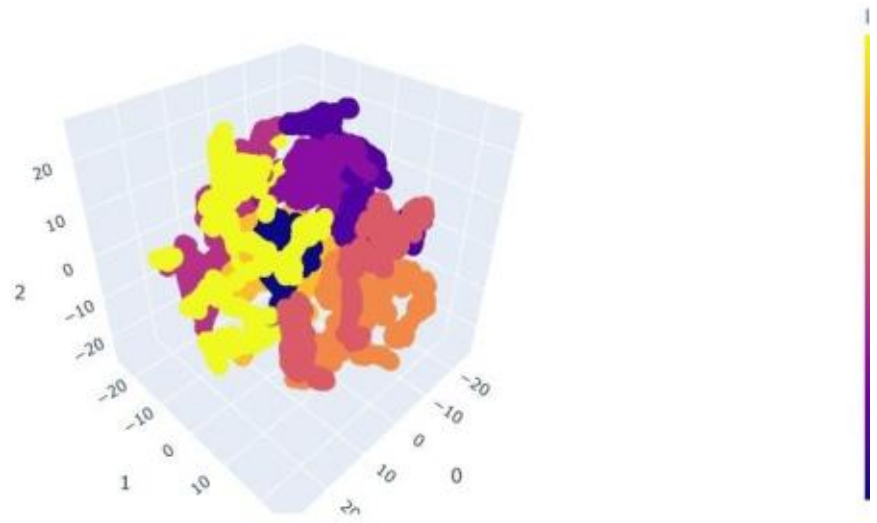

(d) Data visualization by t-SNE 3D

Figure 7. Data visualization by PCA and t-SNE.

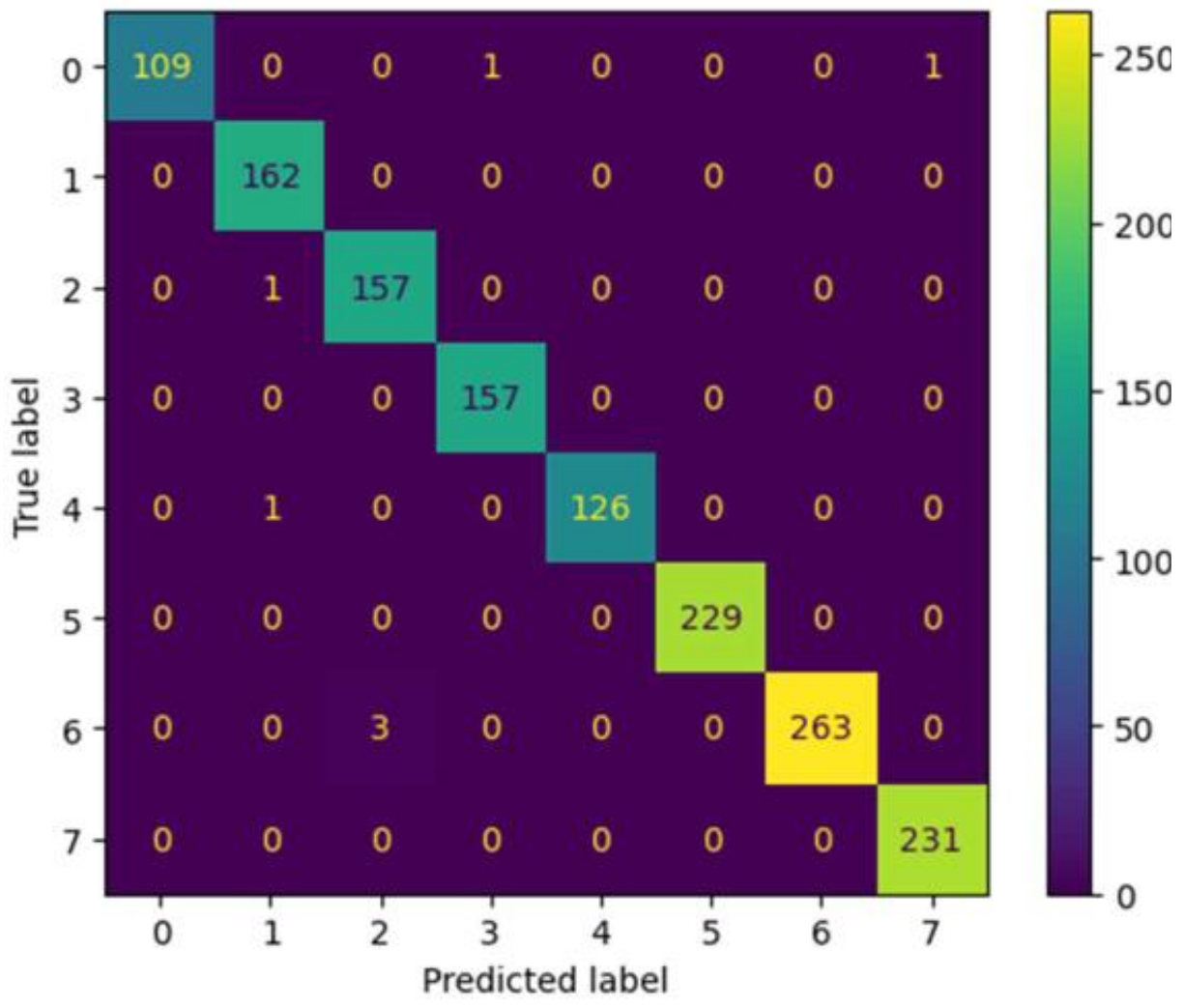

(a) Confusion matrix of the ANN model

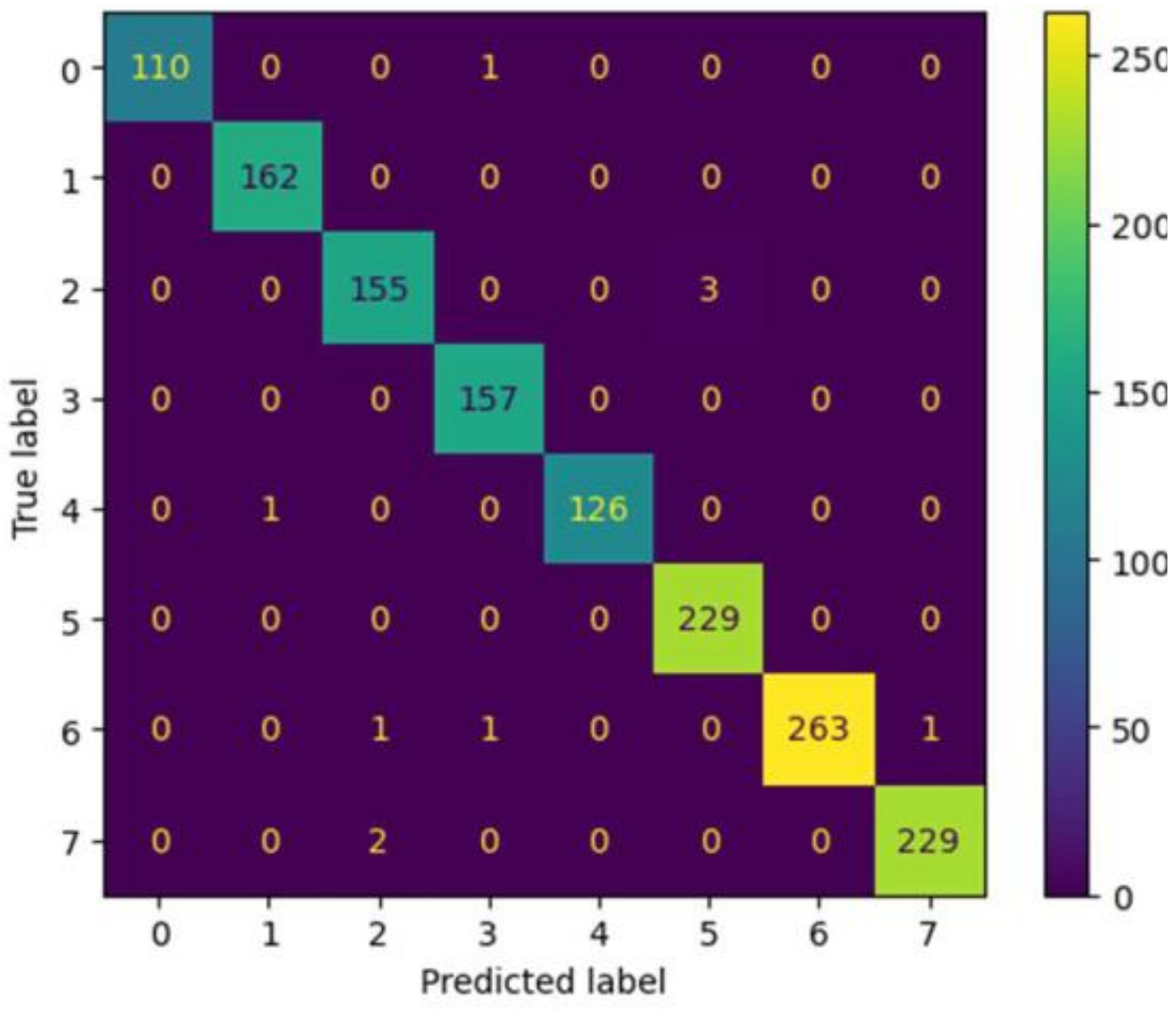

(b) Confusion matrix of the SVM model

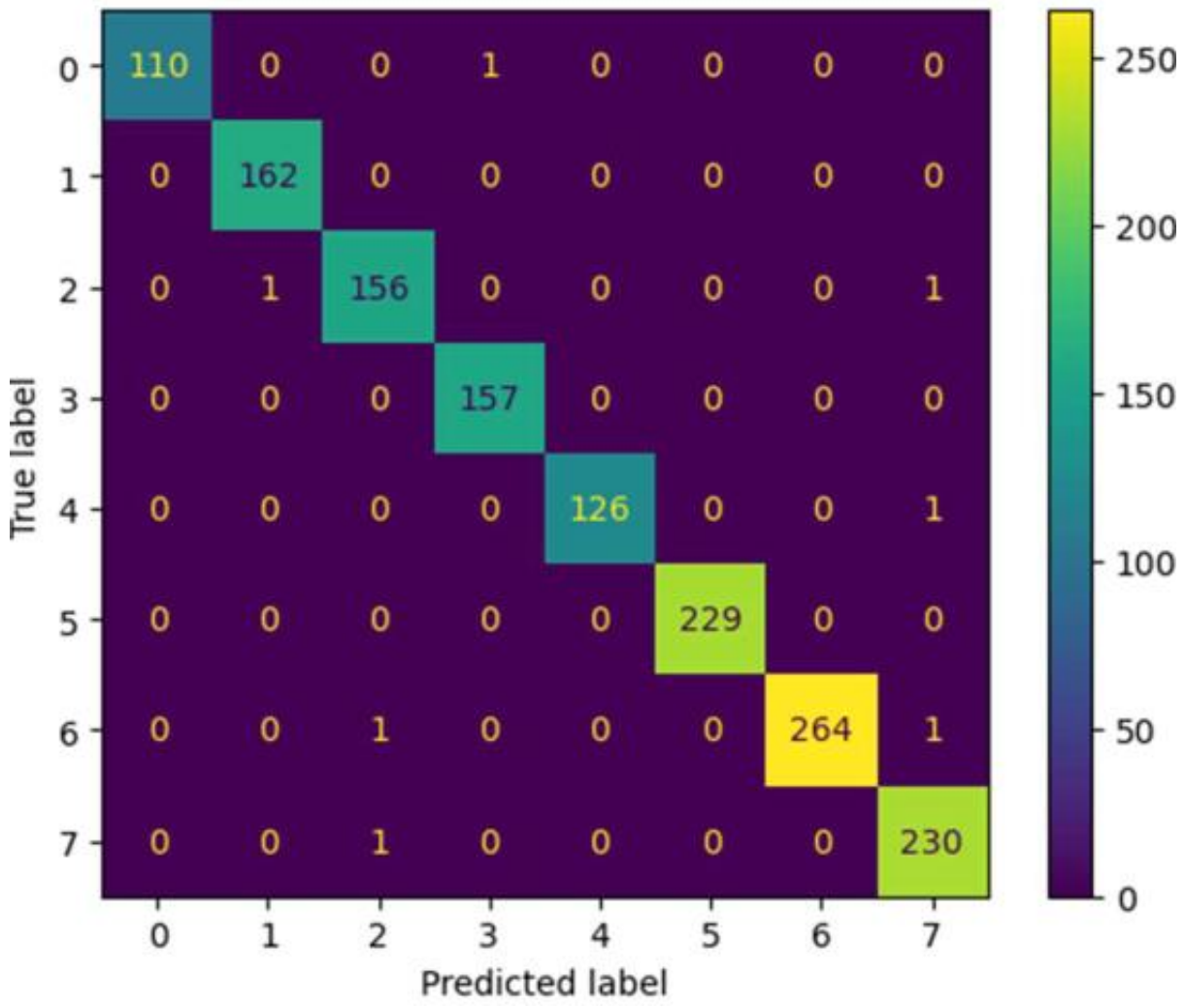

(c) Confusion matrix of the DT model

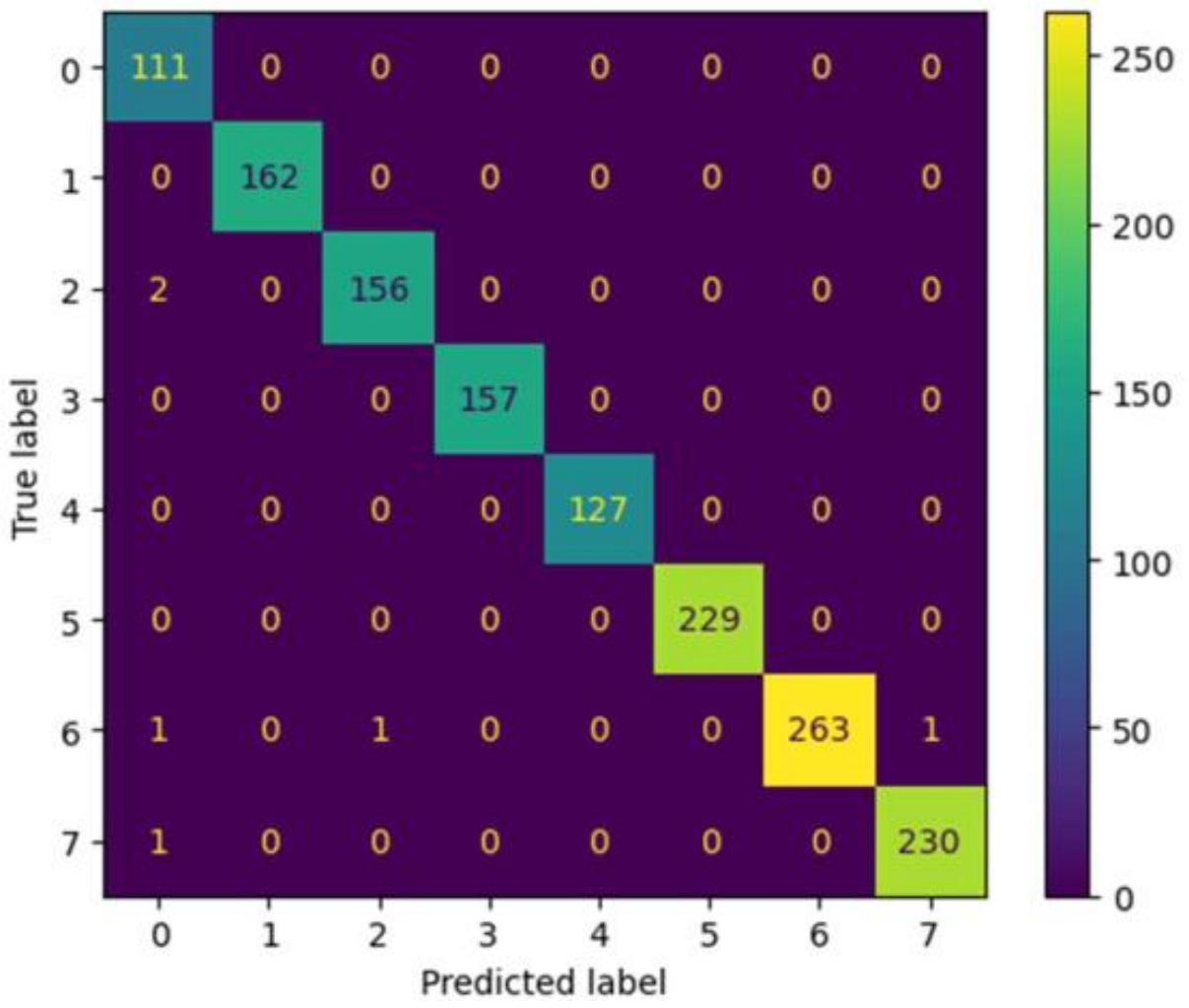

(d ) Confusion matrix of the RF model

Figure 8. Confusion matrix of four models, (a) Confusion matrix of the ANN model, (b) Confusion matrix of the SVM model, (c) Confusion matrix of the DT model, (d) Confusion matrix of the RF model.

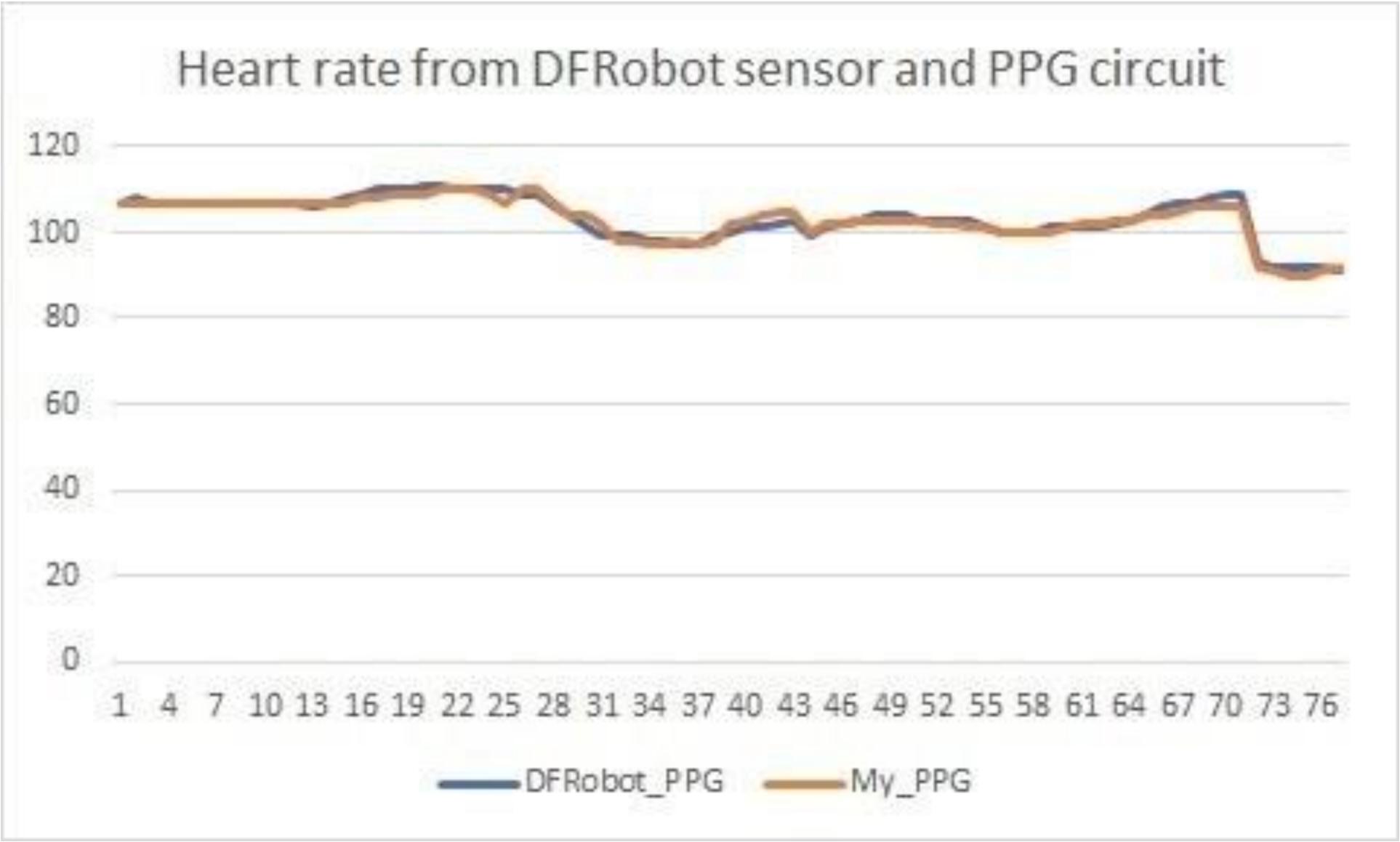

Figure 9. The heart rate from two modules, the DFRobot sensor and the PPG circuit

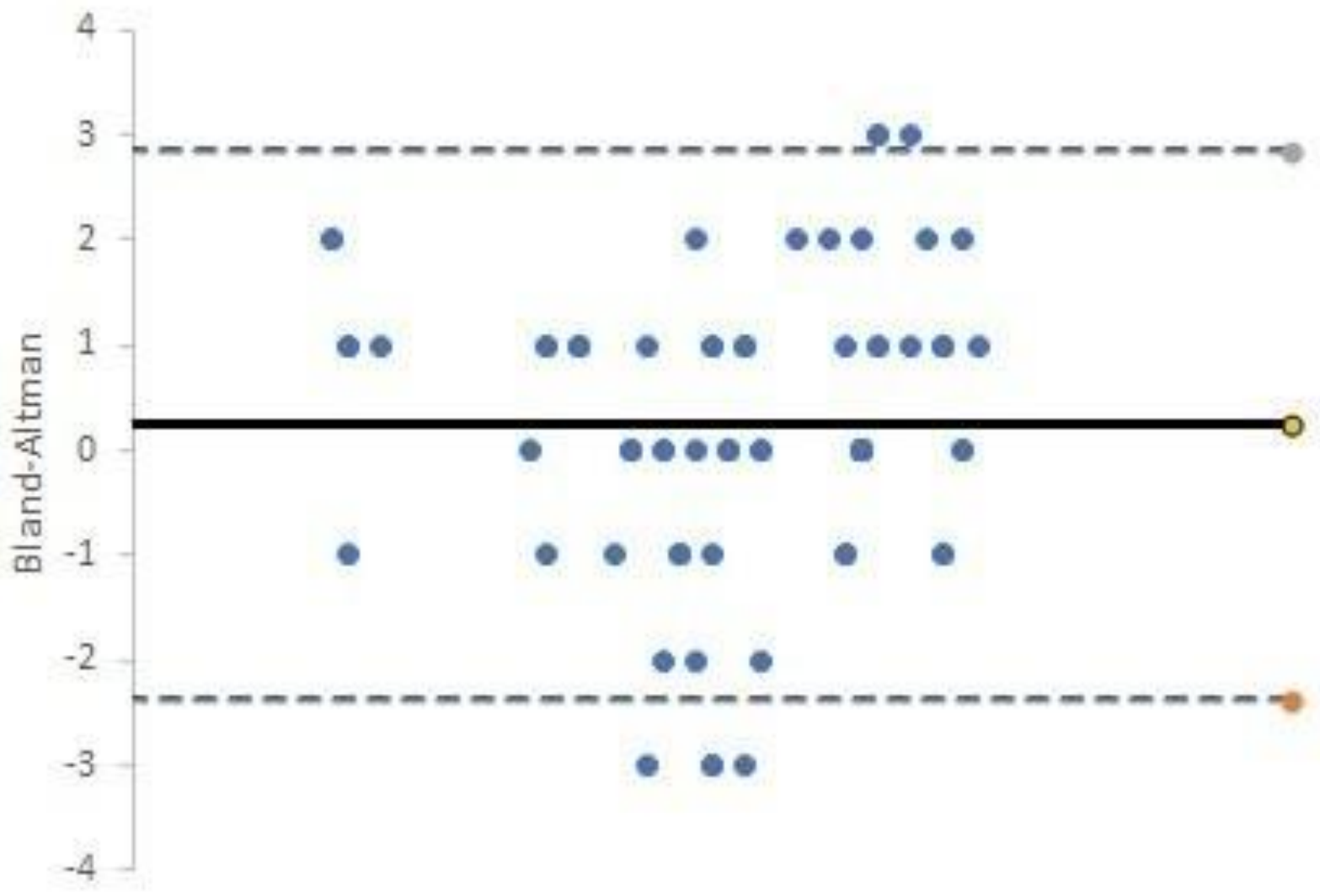

Figure 10. A bland-Altman plot of the DFRobot sensor and the PPG circuit.

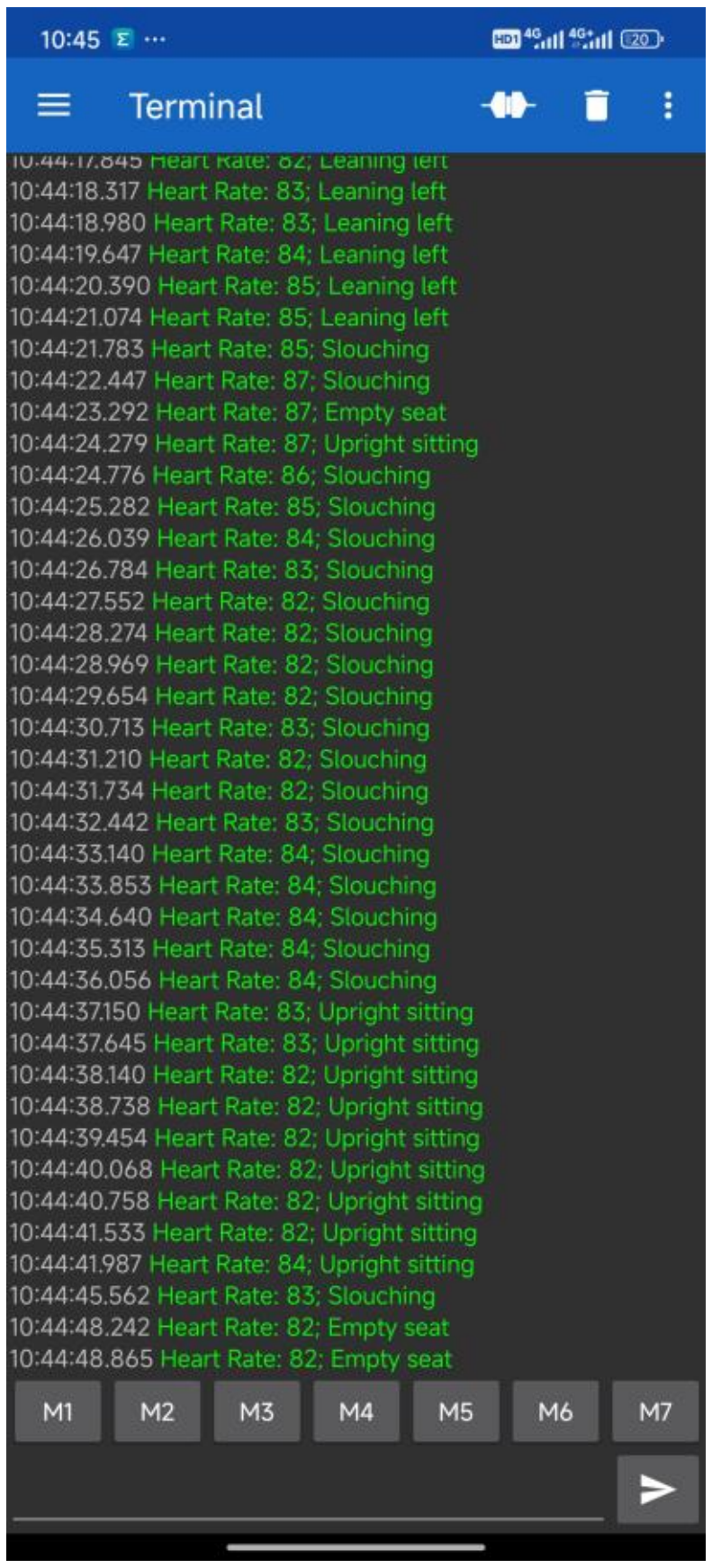

(a) The sitting postures and the heart rate are continuously displayed on the phone.

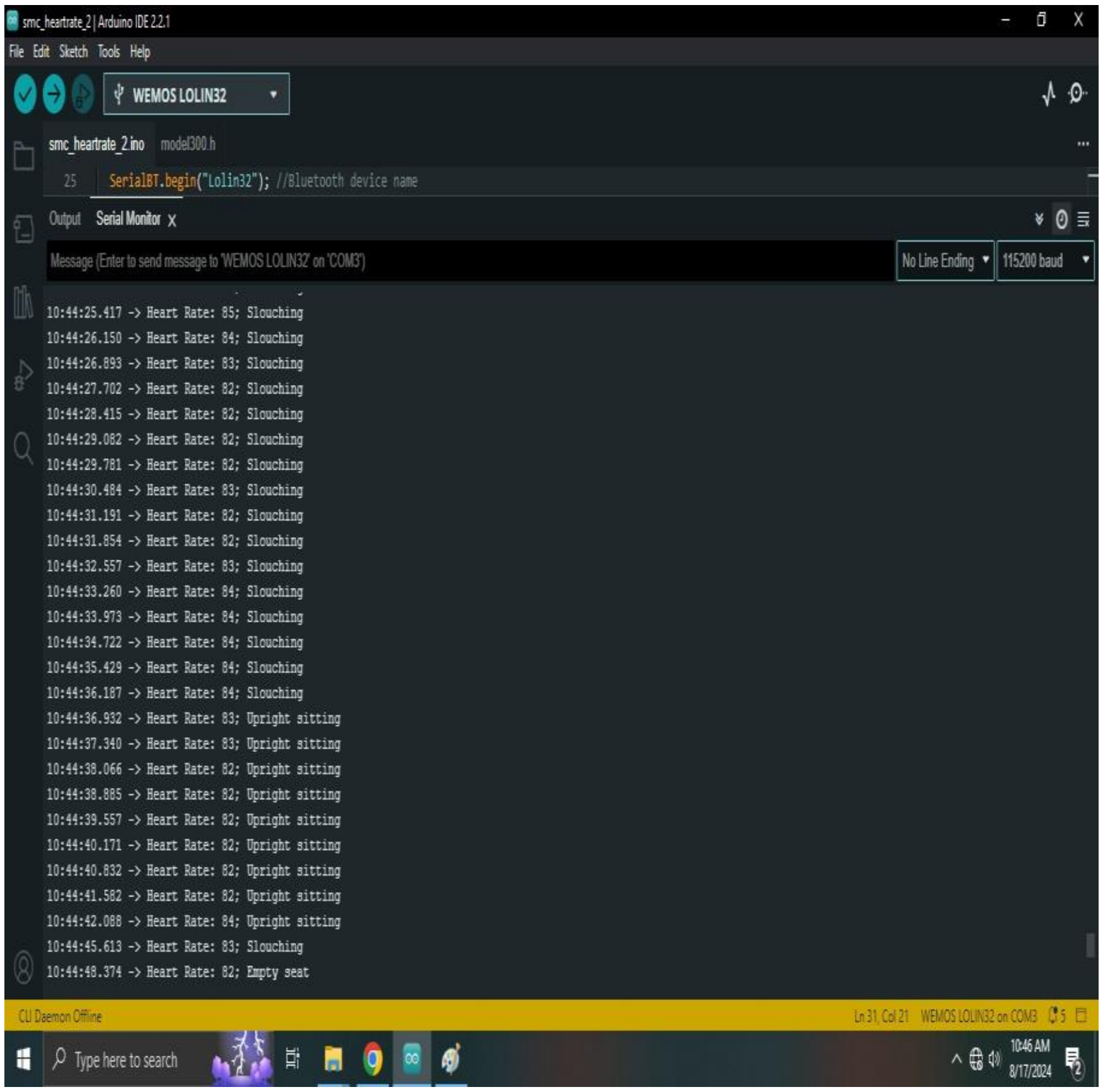

(b) The sitting postures and the heart rate are continuously displayed on the laptops

Figure 11. The sitting postures and the heart rate on (a) the mobile phone and (b) the laptop

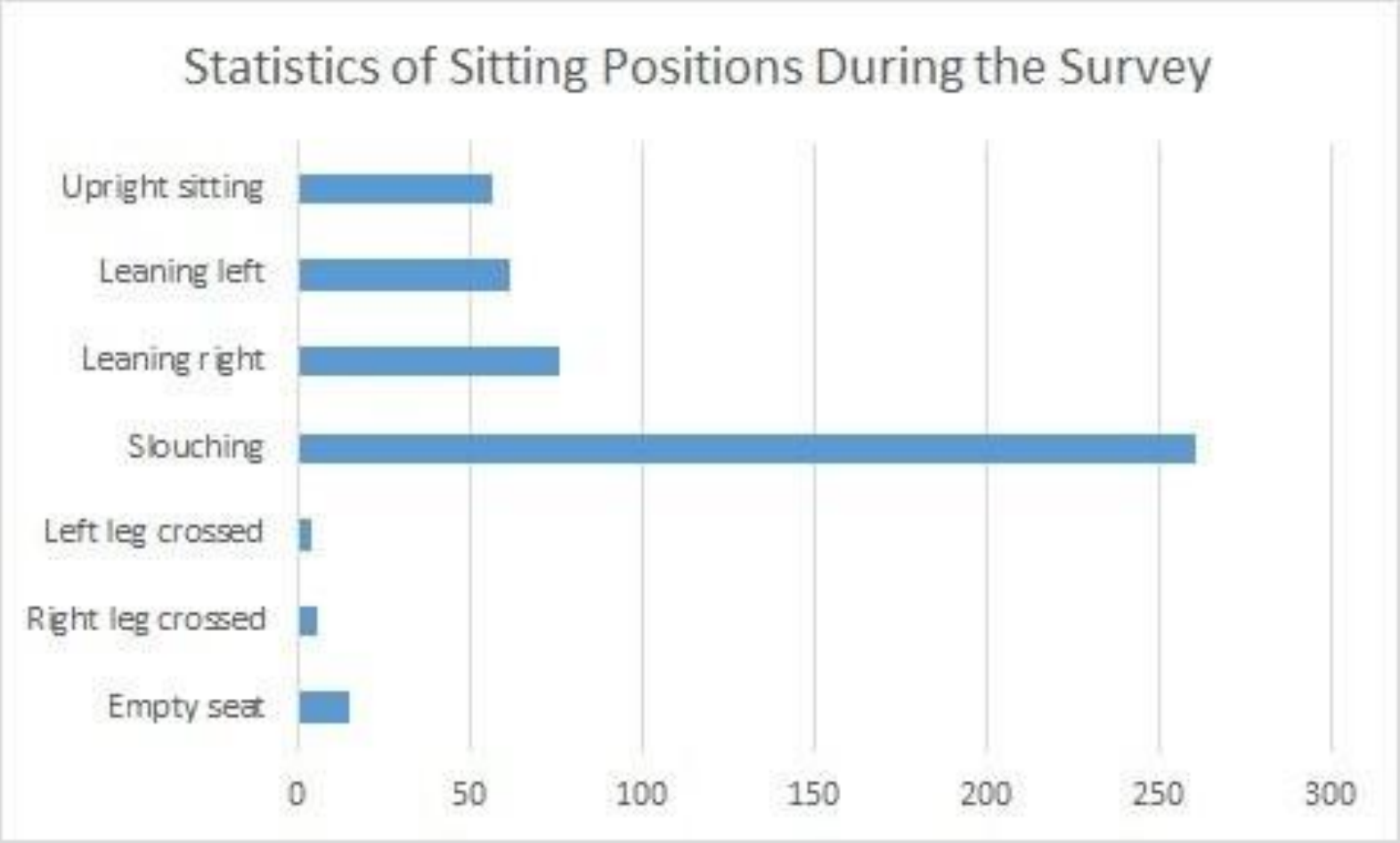

Figure 12. Number of repetitions of sitting postures of a volunteer in seven minutes.

**Table 1.** Comparison of our work and some studies using a pressure array

| Years | Authors | Sensors | Features | No. of posture | Accuracy |
|---|---|---|---|---|---|
| 2013 | Wenyao Xu et al. [32] | Textile Pressure Sensor (16x16 sensors) | Sitting pressure distribution | 7 | 85% |
| 2017 | Mengjie Huang et al. [43] | Piezo-Resistive Sensor (52 × 44) | The pressure patterns | 8 | 92% |
| 2021 | Qilong Wan et al. [44] | Pressure Sensor (32x32) | Sitting pressure image | 4 | 89.60% |
| 2022 | Zhe Fan et al. [49] | Pressure Sensor (44×52) | Pressure heat map | 5 | 99.82% |
| | Our work | Ten sensors | Pressure map | 7 | 99.3% |

**Table 2.** Comparison of our work and some studies placing sensors in other places except for seat cushions.

| Years | Authors | Sensors | Method | No. of posture | Accuracy |
|---|---|---|---|---|---|
| 2022 | Taraneh Aminosharieh Najafi et al. [57] | 8 Force Sensing Resistors (3 sensors on the backrest and 5 sensors on the seat) | - | 8 | 91% |
| 2020 | Haeseok Jeong et al. [56] | 6 Pressure Sensors on the seat and 6 Infrared Reflective Distance Sensors on the backrest | Pressure distribution and seatback-trunk distances | 11 | 92% |
| 2016 | Roland Zemp et al. [39] | Ten pressure sensors on seat, four on the backrest, and three on each armrest | Sixteen force sensor values and the backrest angle | 7 | 90% |
| 2010 | Meyer J. et al. [30] | 96 sensors on the seat pan and one sensor on the backrest | - Sensor value from each sensor element.<br>- Center of force.<br>- Pressure applied to 4 and 16 equal aggregated areas of the seating area | 16 | 82% |
| 2001 | Tan et al. [58] | Sensor arrays (42 × 48) on a chair's seat and backrest (42 × 48) | 2-D grayscale image | 14 | 96% and 79% for familiar and unfamiliar |
| | | | | | users, respectively |
| 2023 | Cátia Tavares et al. [37] | 4 FSR Sensors on the seat and 4 Load Cells on the backrest | Pressure values | 6 | 100% |
| | Our work | Ten sensors | Pressure map | 7 | 99.3% |